\documentclass[11pt]{article}
\usepackage[utf8]{inputenc}
\usepackage[a4paper]{geometry}
\usepackage{amsmath}
\usepackage{geometry}
\usepackage{amsmath}
\usepackage{comment}
\usepackage{amsfonts}
\usepackage{amssymb}
\usepackage{ragged2e}
\newcommand{\sym}[1]{\textsuperscript{#1}}
\usepackage{xcolor}
\usepackage{graphicx}
\usepackage{booktabs}
\usepackage{tabularx}
\usepackage[table]{xcolor}
\usepackage{colortbl}
\usepackage{placeins}
\usepackage{threeparttable}
\usepackage{booktabs}
\usepackage{float}
\usepackage[hidelinks]{hyperref}
\usepackage{setspace}
\usepackage{mathtools}
\usepackage[skip=0pt]{caption}
\usepackage{subcaption}
\usepackage[bottom]{footmisc}
\usepackage{float}
\usepackage{authblk}
\usepackage[backend=biber, style=authoryear, sorting=nyt, uniquename = false]{biblatex}
\makeatletter

\title{Are Princelings Truly Busted? Evaluating Transaction Discounts in China's Land Market}
\author{Julia Manso\thanks{Correspondence: Julia Manso, Department of Statistics, 24-29 St Giles', Oxford OX1 3LB, United Kingdom. \newline Email: jumanso@stats.ox.ac.uk. I thank Frank Windmeijer for his advice, guidance, and feedback throughout the drafting of this paper. }}
\affil{\small{\textit{Department of Statistics and Nuffield College, University of Oxford, Oxford, U.K.}}}
\date{}

\begin{document}

\maketitle
\singlespacing
\vspace{-10mm}
\begin{abstract}
    This paper replicates Chen and Kung's 2019 analysis (\textit{The Quarterly Journal of Economics} 134(1): 185-226). Inspecting the data reveals that nearly one-third of transactions (388,903 out of 1,208,621) are perfect duplicates of other rows, excluding the transaction number. The analysis on the data sans duplicates replicates their statistically significant princeling effect, robust across various specifications. Further analysis reveals a disagreement between Chen and Kung's text and code: the paper's ``logarithm of area" is actually area ($\text{m}^2$) divided by one million. This therefore necessitates a reinterpretation of the estimation results, revealing that the princeling effect is extremely large.
\end{abstract}
\noindent \textbf{Keywords\textemdash} data reproducibility, empirical validation, institutional reform, Chinese real estate

\vspace{-7mm}
\doublespacing
\section{Introduction}
\vspace{-2mm}
The recent implosion of several leading Chinese real estate developers, such as Evergrande and Country Garden, has kindled interest in the mechanisms of the Chinese real estate market. Unique not only for its highly leveraged financing practices, the market is also a nexus of interactions between state-owned and private enterprises and has historically been plagued by corruption and cronyism, particularly on the land sale side. One often-cited paper on the topic is Chen and Kung (2019), titled ``Busting the `Princelings': The Campaign against Corruption in China's Primary Land Market," which explores the unique mechanism of Chinese cronyism in land sale. Specifically, they examine how ``princeling" firms\textemdash those firms tied to family members of China's top governing body, the Politburo\textemdash receive substantial price discounts on land parcels after controlling for several location-based and transaction-level variables (\cite{chen_busting_2019}). They also have significant findings relating princeling power, promotion likelihood, and magnitude of discount: more powerful princelings obtain larger discounts, and provincial party secretaries providing discounts were more likely to be promoted, with the likelihood increasing by the magnitude of the discount and the area of discounted land sold. 

Between containing detailed information about officials' promotions and establishing firms' state-ownership and princeling statuses, the paper's replication data is also of popular interest and has over 12,200 downloads from its archive on the Harvard Dataverse.\footnote{The paper's replication data can be found at \href{https://doi.org/10.7910/DVN/XW6OJT}{\texttt{https://doi.org/10.7910/DVN/XW6OJT}}.} 

This paper replicates the results of Chen and Kung (2019, hereafter ``CK"). A previous investigation by \textcite{wiebe_replicating_2024} diagnosed issues with CK's variable on mayoral promotion. Yet, Wiebe found that the result was unchanged even when a corrected promotion variable was used. My replication identifies more extensive issues: an initial inspection of CK's data reveals that nearly one-third of the transactions (388,903 out of 1,208,621) are perfect duplicates of other rows, excluding the transaction number. These duplicates suggest, for instance, that the same buyer purchased two (or more) identical parcels of land with the same quality rating, area, and usage in the same city, month, and year, at the same price. Examining records of land sale transactions suggests that some of these duplicates correctly refer to identical yet distinct parcels sold at the same time, but many other duplicates erroneously reference the same parcel of land multiple times. Importantly, over one-half of the princeling parcels in the full dataset (9,815 out of 19,812) are duplicates. Replicating the analysis on the data, sans duplicates, yields similar results to CK's original findings: while the discount magnitude of princelings is slightly smaller than that found using the original dataset, the princeling effect is still large and statistically significant across the regression results. 

Yet, further inspection reveals that when CK aggregate parcel areas to group parcels (for instance, summing transaction data to calculate the total area of land sold to each firm in each year), their text and code differ: the ``logarithm of area" in CK's paper is actually area ($\text{m}^2$) divided by 1,000,000. I thus reinterpret CK's originally calculated coefficients as scaled area values rather than logged area values, with and without duplicates. On this scale, the magnitude of the princeling effect is quite large, suggesting that princelings on average purchase tens of thousands of additional square meters of land per year compared to their nonprinceling peers\textemdash areas double, triple, or even quadruple the median parcel size, depending on the specification. In the  appendix, I include calculations utilizing the logarithm of area purchased; these results are strange and somewhat unstable, with the princeling effect not only changing sign and direction between tables, but also dropping sharply prior to the anticorruption campaign. 
\vspace{-5mm}

\section{Apparent Duplicates in the Data} \label{Apparent Duplicates in the Data}
\vspace{-2mm}
In their paper, CK construct a set of ``princeling" firms from several online sources and match them to land transaction data obtained from the Ministry of Land and Resources. I focus on this latter data, extracted from the Land Transaction Monitoring System (\href{http://www.landchina.com/}{\texttt{www.landchina.com}}). As each municipality's ``bureau of land and resources is required to report each land transaction in their jurisdiction electronically on this website," per the Law of Land Management (p. 199), the dataset captures all land transactions.\footnote{China eliminated the secondary market for land use rights (LURs) in 2004, meaning the State is the sole seller of land, doing so at the municipal level (\cite{li_urban_2019}; \cite{gyourko_land_2022}).} The Land Transaction Monitoring System provides comprehensive information on land transactions, relaying everything from transaction details like total payment amount, sale method, buyer name, and total area of land sold, to building restrictions, like floor area ratio and greening ratio, to precise location details on the parcel.\footnote{CK note that the system also provides the address and postal code of the parcel, ``a two-digit code of land usage... [and] a three-digit industry code of the buyer's firm" (p. 200). These fields are no longer included on the Land Transaction Monitoring System website nor are they included in CK's replication data, which makes determining the extent of duplicates more complicated; it is possible that these information fields were removed from the database sometime after CK collected the data. Alternatively, CK may have obtained a slightly more comprehensive version of the website's data directly from the Ministry.} 

Examining CK's dataset reveals several cases of apparent duplicates: specifically, after removing the transaction identifier, which was likely added by CK in data processing, 388,903 land transactions appear to be perfect duplicates, suggesting that a single buyer purchased identical parcels in the same month, year, and city for the same price and use. Table \ref{tab:table_of_duplicates} shows the extent and distribution of these duplicates. Of the 1,208,621 land transactions in CK's data,  819,718 are unique, and 299,167 of these have at least one duplicate. While most transactions have only one duplicate, others have several\textemdash as many as 158. Importantly, of the 19,812 transactions that were conducted by princelings in the original data, over half of these are duplicates, with a significant share having three or more seemingly identical transactions. Duplicates are especially prevalent after 2012\textemdash the year that the anti-corruption campaign began. 

\vspace{4mm}
\begin{table}[!h]
\centering
\caption{Summary of duplicates by frequency of occurrence and type}
\centering
\resizebox{\textwidth}{!}{
\begin{tabular}[t]{lrrrrr}
\toprule
\multicolumn{1}{c}{ } & \multicolumn{3}{c}{Total Observations} & \multicolumn{2}{c}{Observations by Time Period} \\
\cmidrule(l{3pt}r{3pt}){2-4} \cmidrule(l{3pt}r{3pt}){5-6}
Frequency of occurrence & All & State-owned & Princeling & 2012 and earlier & After 2012\\
\midrule
1 & 520,551 & 181,571 & 4,414 & 253,322 & 267,120\\
2 & 218,001 & 75,640 & 3,479 & 191,718 & 26,283\\
3 & 79,366 & 27,473 & 1,554 & 79,200 & 166\\
4 & 581 & 193 & 81 & 518 & 63\\
5-10 & 966 & 319 & 201 & 870 & 96\\
11-20 & 167 & 59 & 48 & 134 & 33\\
21-30 & 48 & 10 & 21 & 34 & 14\\
31-158 & 38 & 15 & 17 & 29 & 9\\
\addlinespace
Number of unique rows & 819,718 & 285,280 & 9,815 & 525,825 & 293,784\\
Number of total rows in Chen and Kung & 1,208,621 & 420,004 & 19,812 & 886,116 & 322,396\\
Number of duplicate rows & 388,903 & 134,724 & 9,997 & 360,291 & 28,612\\
Number of unique rows with duplicates & 299,167 & 103,709 & 5,401 & 272,503 & 26,664\\
\bottomrule
\label{tab:table_of_duplicates}
\end{tabular}
}
\end{table}
\vspace{-2mm}

In light of these results, there are a couple of possibilities: first, that there are truly identical neighboring parcels that are sold at the same time and price to a single buyer for specific usages. Second, it is possible that the duplicate rows are erroneous and that the database itself has multiple land transactions that refer to the same underlying land parcel. Examining CK's replication data is not enough to confirm either possibility, as the data do not include each parcel's specific address. Yet, the transaction details can be matched with the original data from the Ministry of Land's database, and this additional information can be used to determine if multiple transaction listings refer to the same parcel. Specifically, after being identified in the database, each land transaction can be linked with a parcel number via the transaction's original sale announcement (the ``Transfer Announcement"), and one can confirm how many parcels of each size are up for sale in the area at the time. With this information, identical land transaction listings that refer to the same underlying parcel can be confirmed as erroneous duplicates. 

Looking at the detailed transaction records indicates that both hypotheses are true: while identical parcels were sometimes sold to a single buyer, many other duplicates refer to one underlying parcel and are thus erroneous. Some of these are quite easy to recognize as duplicate listings of the same land transaction when referencing the Ministry of Land's database, as they may even have the same electronic lookup number\textemdash a ``unique" identifier\textemdash for multiple transactions.\footnote{The electronic lookup number is the database's main identifier/search key for unique land transactions. Having two ``separate" transactions with the same electronic lookup number is thus impossible and indicates an error.} The Transfer Announcements also confirm there should only be one listing for the parcel in the Ministry of Land's database\textemdash and in CK's data. 

Other cases are less clear but strongly indicative of incorrect duplicates. For instance, two transactions have different electronic lookup numbers, but the Transfer Announcement suggests that both listings refer to the same underlying parcel.\footnote{In many cases, examining the full transaction details of the two listings reveals missing information in one listing. These omissions are non-essential details (e.g., greening ratio, floor area ratio) rather than key identifiers (e.g., buyer, area, quality). A second listing of the same parcel may thus exist simply to complete the initial record.} Cases with extremely high numbers of duplicates appear to largely be erroneous; for instance, for one entry with 71 apparent duplicates in CK's data, there is no record of even one transaction occurring with the given parcel features.

Additional cases appear to be partially correct, partially incorrect duplicates. For instance, the Ministry of Land's database reveals two identical parcels were sold at the same time to a single buyer at the same price. Yet, CK's dataset has three transactions that match these features rather than two.\footnote{This may reflect an erroneous duplicate entry being removed from the website after CK obtained their data.} A sample of several such cases is included in the  appendix. 

Further, some transactions appear identical except for the location dummy variables (\texttt{near500} and \texttt{near1500}) created by CK. These are dummy variables that reflect whether a given parcel is within 500 or 1500 meters, respectively, of a parcel purchased by a princeling firm. Presumably, CK used the specific location data (not included in their replication files) to determine these binary values, and thus, having different binaries for these values implies that the parcels are indeed distinct. These rows are therefore not classified as duplicates in this paper.\footnote{There are approximately 100,000 of these rows in the data. I was unable to verify whether these parcels are distinct using the Ministry of Land's database alone; the specific address information would be needed to ensure these transactions reflect separate parcels.     Even if one wanted to remove these rows as duplicates, it is not apparent which observation contains the correct \texttt{near500} and \texttt{near1500} values without the parcel addresses. I thus leave these quasi-duplicates in, but note that their status as unique parcels should be verified further. I also conduct robustness checks that exclude these duplicates, as discussed in footnote 10.}

Thus, a matched comparison with the Ministry of Land's database, tracing parcel features and Transfer Announcements, suggests a significant portion of the duplicate rows erroneously refer to the same underlying transaction. Yet, without parcel addresses, it is not possible to examine the exact extent of incorrect duplicates versus those that refer to different but identical parcels.
\vspace{-5mm}

\subsection{Replicating results without duplicates: Tables V and X}
\vspace{-1mm}
When replicating CK's regression results, I drop all duplicate rows but recognize that this path excludes duplicate entries that refer to different but identical parcels. However, as highlighted 
\noindent \makebox[\linewidth][s]{in Table \ref{tab: table 5 unique}, the regression results are largely still similar for their main specification (CK's Table V),} 
\vspace{-6mm}
\begingroup
  \setlength{\abovedisplayskip}{0pt}
  \setlength{\belowdisplayskip}{0pt}
  \begin{equation}
    Price_{ickst} = \beta_0 + \beta_1 PrincelingFirms_{ikt} + \gamma X_i + T_{cst} + \nu_{ickst}.
\end{equation}
\endgroup
\noindent As detailed by CK, the dependent variable $Price_{ickst}$ ``is the log of price (yuan per square meter) for land parcel $i$ sold by local government $c$ to firm $k$ for usage $s$ in month-year $t$. The key explanatory variable, $PrincelingFirms_{ikt}$, is a dummy variable equal to 1 if a firm $k$ is connected to a princeling" and 0 otherwise (p. 203). They control for several transaction-level variables including land quality (as evaluated by the municipal government, on a scale of 1-20), state-ownership status, sale method, firm size, and log(area/100), where area is the land area sold in square meters. They also have city-year-usage, industry, and month fixed effects, with standard errors clustered at the province and firm levels and singletons included.

Column (1) illustrates regression (1) run on the full dataset, while columns (2) and (3) are conducted on the matched sample of parcels within a specified radius (1,500 or 500 meters, respectively) from a princeling parcel. Columns (4), (5), and (6) repeat these regressions but now include princeling firms interacted with the dummy variable $PSCM$ (Politburo Standing Committee Member), aiming to explore whether firms connected to these committee members received larger discounts. Likewise, columns (7), (8), and (9) include princeling firms interacted with a dummy variable that reflects whether a transaction occurred after the connected official was retired. In all of these cases, the princeling effect is strongly negative with small standard errors\textemdash as was the case in CK's original results\textemdash suggesting that the princeling effect is robust despite the apparent duplicates. Interestingly, when duplicates are removed, the princeling effect itself appears slightly smaller; for instance, the price advantage for princeling-connected firms, based on results for the 500-meter-radius sample, is now a 55.3\% (($1-\text{exp}(-0.804))\times100$) discount compared to CK's original 57.0\% (($1-\text{exp}(-0.844))\times100$) discount. However, the interacted terms (princeling firms with Politburo Standing Committee connections and princeling firms connected to retired officials) are consistently larger than the original data's results.\footnote{As a robustness check, I also replicate these main regression specifications using a stronger definition of duplicates\textemdash that is, when defining a unique row, I exclude all variables added in data processing, like \texttt{near500} and \texttt{near1500}. Under this definition, there are 727,127 unique rows. The regression results shrink slightly more towards zero but remain largely the same. Standard errors likewise remain similar.}

\vspace{3mm}
\setlength{\textfloatsep}{0pt}
\setlength{\tabcolsep}{1pt} 
\begin{table}[H] 
\centering
\caption{CK's Table V}
\label{tab: table 5 unique}
\resizebox{\textwidth}{!}{
\begin{threeparttable}
\setlength{\extrarowheight}{2pt} 
\begin{tabular}{l*{9}{c}}
\hline \hline
                    & \multicolumn{9}{c}{Log of land price} \\
\cmidrule(lr){2-10}
                    & All
                    & \multicolumn{1}{c}{\shortstack{$\leq 1,500$\\meters}} 
                    & \multicolumn{1}{c}{\shortstack{$\leq 500$\\meters}}
                    & All 
                    & \multicolumn{1}{c}{\shortstack{$\leq 1,500$\\meters}} 
                    & \multicolumn{1}{c}{\shortstack{$\leq 500$\\meters}}
                    & All 
                    & \multicolumn{1}{c}{\shortstack{$\leq 1,500$\\meters}} 
                    & \multicolumn{1}{c}{\shortstack{$\leq 500$\\meters}} \\
                    & (1) & (2) & (3) & (4) & (5) & (6) & (7) & (8) & (9) \\
\midrule
Original data \\
\quad Princeling firms    &      -0.808\sym{***}&      -0.904\sym{***}&      -0.844\sym{***}&      -0.545\sym{***}&      -0.666\sym{***}&      -0.620\sym{***}&      -0.808\sym{***}&      -0.894\sym{***}&      -0.835\sym{***}\\
                    &     (0.025)         &     (0.034)         &     (0.033)         &     (0.035)         &     (0.043)         &     (0.043)         &     (0.030)         &     (0.040)         &     (0.038)         \\
\quad Princeling firms*PSCM&                     &                     &                     &      -0.442\sym{***}&      -0.420\sym{***}&      -0.396\sym{***}&                     &                     &                     \\
                    &                     &                     &                     &     (0.037)         &     (0.048)         &     (0.049)         &                     &                     &                     \\
\quad Princeling firms*Retired&                     &                     &                     &                     &                     &                     &      -0.001         &      -0.051         &      -0.044         \\
                    &                     &                     &                     &                     &                     &                     &     (0.056)         &     (0.063)         &     (0.058)         \\
\cmidrule(lr){2-10}
\quad Number of observations & 1,144,507 & 334,232 & 191,896 & 1,144,507 & 334,232 & 191,896 & 1,144,507 & 334,232 & 191,896 \\
\quad Adjusted \(R^{2}\)     & 0.692 & 0.727 & 0.755 & 0.692 & 0.728 & 0.756 & 0.692 & 0.727 & 0.755 \\
\midrule 
Data without duplicates \\ 
\quad Princeling firms    &      -0.750\sym{***}&      -0.849\sym{***}&      -0.804\sym{***}&      -0.485\sym{***}&      -0.602\sym{***}&      -0.574\sym{***}&      -0.737\sym{***}&      -0.830\sym{***}&      -0.789\sym{***}\\
                    &     (0.027)         &     (0.033)         &     (0.033)         &     (0.035)         &     (0.039)         &     (0.041)         &     (0.033)         &     (0.038)         &     (0.037)         \\
\quad Princeling firms*PSCM&                     &                     &                     &      -0.488\sym{***}&      -0.470\sym{***}&      -0.436\sym{***}&                     &                     &                     \\
                    &                     &                     &                     &     (0.040)         &     (0.048)         &     (0.048)         &                     &                     &                     \\
\quad Princeling firms*Retired&                     &                     &                     &                     &                     &                     &      -0.063         &      -0.090         &      -0.069         \\
                    &                     &                     &                     &                     &                     &                     &     (0.057)         &     (0.062)         &     (0.056)         \\
                    \cmidrule(lr){2-10}
\quad Number of observations  &      779,372         &      248,627         &      136,480         &      779,372         &      248,627         &      136,480         &      779,372         &      248,627         &      136,480         \\
\quad Adjusted \(R^{2}\)  &       0.693         &       0.716         &       0.736         &       0.694         &       0.717         &       0.737         &       0.693         &       0.716         &       0.736         \\
\hline \hline
\end{tabular}
\end{threeparttable}
}
\caption*{\footnotesize Note: Following CK, all columns include city-year-usage, month, and industry fixed effects, as well as the following control variables: land quality (on a scale of 1-20), the logarithm of land area sold, firm size (on a scale of 0-3), state-ownership status, and dummy variables for sale method. Standard errors are robust to two-way clustering by firm and province and are in parentheses (\sym{*} \(p<0.05\), \sym{**} \(p<0.01\), \sym{***} \(p<0.001\)).
}
\end{table}

\vspace{-3mm}

CK also use this granular transaction data in their Table X, ``Princeling Purchases and Land Prices after Xi Took Office," which examines whether President Xi's anticorruption campaign impacted the price discount given to princeling firms. I regenerate this table without duplicate rows (Table \ref{tab: Table 10 no dups}), and the results are again quite similar to CK's original findings: for the 500-meter matched sample, CK found that princeling firms had previously obtained an average price discount of 60.1\% (($1-\text{exp}(-0.920))\times100$) that dropped by 11.7\% (($1-\text{exp}(-0.920) - [1-\text{exp}(-0.920 +0.257)])\times100$) after the introduction of President Xi's anticorruption campaign (column (2)); without duplicates, the 60.2\% (($1-\text{exp}(-0.923))\times100$) discount before the anticorruption campaign drops by 13.8\% (($1-\text{exp}(-0.923) - [1-\text{exp}(-0.923 +0.298)])\times100$) after the campaign begins (Table \ref{tab: Table 10 no dups}, column (2)). The other columns are equally similar and again suggest the anticorruption campaign strongly impacts the magnitude of discounts obtained by princelings; these remaining columns explore changes in princeling advantage in the presence of interacted dummy variables involving different dimensions of the anticorruption campaign (e.g., 

\vspace{3mm}
\setlength{\tabcolsep}{1pt}
\begin{table}[H] 
\centering
\caption{CK's Table X}
\label{tab: Table 10 no dups}
\resizebox{\textwidth}{!}{
\begin{threeparttable}
\setlength{\extrarowheight}{2pt} 
\begin{tabular}{l*{11}{c}} 
\hline \hline
                    & \multicolumn{10}{c}{Log of land price} \\ 
\cmidrule(lr){2-11}
                    & All 
                    & \multicolumn{1}{c}{\shortstack{$\leq500$\\meters}} 
                    & All
                    & \multicolumn{1}{c}{\shortstack{$\leq500$\\meters}}  
                    & All
                    & \multicolumn{1}{c}{\shortstack{$\leq500$\\meters}}
                    & All
                    & \multicolumn{1}{c}{\shortstack{$\leq500$\\meters}}
                    & All
                    & \multicolumn{1}{c}{\shortstack{$\leq500$\\meters}} \\
                    & (1) & (2) & (3) & (4) & (5) & (6) & (7) & (8) & (9) & (10) \\ 
\midrule
Original data \\
\quad Princeling firms    & -0.907\sym{***} & -0.920\sym{***} & -0.825\sym{***} & -0.858\sym{***} & -0.870\sym{***} & -0.896\sym{***} 
                    & -0.907\sym{***} & -0.920\sym{***} & -0.818\sym{***} & -0.847\sym{***} \\
                    & (0.029) & (0.040) & (0.024) & (0.032) & (0.028) & (0.035) 
                    & (0.029) & (0.040) & (0.023) & (0.028) \\
\quad Princeling firms     & 0.318\sym{***} & 0.257\sym{***} &  &  &  &  & 0.140\sym{*} & 0.093 &  &  \\
\hspace{15pt} $*$Transaction after 2012 & (0.047) & (0.058) &  &  &  &  & (0.052) & (0.054) &  &  \\
\quad Princeling firms     &  &  & 0.819\sym{***} & 0.695\sym{***} &  &  & 0.504\sym{***} & 0.420\sym{***} &  &  \\ 
\hspace{15pt} $*$Central inspection &  &  & (0.124) & (0.139) &  &  & (0.079) & (0.096) &  &  \\
\quad Princeling firms     &  &  &  &  & 0.614\sym{***} & 0.572\sym{***} & 0.449\sym{***} & 0.447\sym{***} &  &  \\ 
\hspace{15pt} $*$Xi-appointed officials &  &  &  &  & (0.055) & (0.051) & (0.064) & (0.059) &  &  \\
\quad Princeling firms     &  &  &  &  &  &  &  &  & 0.109 & 0.037 \\
\hspace{15pt} $*$Pre-2012 inspection &  &  &  &  &  &  &  &  & (0.074) & (0.070) \\
\cmidrule(lr){2-11}
\quad Number of observations & 1,144,507 & 191,896 & 1,144,507 & 191,896 & 1,144,507 & 191,896 & 1,144,507 & 191,896 & 1,144,507 & 191,896 \\
\quad Adjusted \(R^{2}\)    & 0.692 & 0.755 & 0.692 & 0.755 & 0.692 & 0.755 & 0.692 & 0.756 & 0.692 & 0.755 \\
\midrule
Data without duplicates \\
\quad Princeling firms    &      -0.899\sym{***}&      -0.923\sym{***}&      -0.774\sym{***}&      -0.823\sym{***}&      -0.844\sym{***}&      -0.884\sym{***}&      -0.900\sym{***}&      -0.923\sym{***}&      -0.752\sym{***}&      -0.802\sym{***}\\
                    &     (0.033)         &     (0.041)         &     (0.025)         &     (0.031)         &     (0.034)         &     (0.037)         &     (0.033)         &     (0.041)         &     (0.025)         &     (0.028)         \\
\quad Princeling firms &       0.351\sym{***}&       0.298\sym{***}&                     &                     &                     &                     &       0.151\sym{**} &       0.113\sym{*}  &                     &                     \\
\hspace{15pt} $*$Transaction after 2012 &     (0.047)         &     (0.059)         &                     &                     &                     &                     &     (0.051)         &     (0.053)         &                     &                     \\
\quad Princeling firms &                     &                     &       0.766\sym{***}&       0.675\sym{***}&                     &                     &       0.488\sym{***}&       0.415\sym{***}&                     &                     \\ 
\hspace{15pt} $*$Central inspection &                     &                     &     (0.115)         &     (0.135)         &                     &                     &     (0.081)         &     (0.099)         &                     &                     \\
\quad Princeling firms &                     &                     &                     &                     &       0.595\sym{***}&       0.571\sym{***}&       0.441\sym{***}&       0.445\sym{***}&                     &                     \\
\hspace{15pt} $*$Xi-appointed officials &                     &                     &                     &                     &     (0.055)         &     (0.054)         &     (0.065)         &     (0.060)         &                     &                     \\
\quad Princeling firms &                     &                     &                     &                     &                     &                     &                     &                     &       0.028         &      -0.025         \\
\hspace{15pt} $*$Pre-2012 inspection &                     &                     &                     &                     &                     &                     &                     &                     &     (0.077)         &     (0.073)         \\
\cmidrule(lr){2-11}
\quad Number of observations & 779,372 & 136,480 & 779,372 & 136,480 & 779,372 & 136,480 & 779,372 & 136,480 & 779,372 & 136,480 \\
\quad Adjusted \(R^{2}\)     & 0.694 & 0.736 & 0.693 & 0.736 & 0.694 & 0.737 & 0.694 & 0.737 & 0.693 & 0.736 \\
\hline \hline
\end{tabular}
\end{threeparttable}
}
\caption*{\footnotesize Note: Following CK, all columns include city-year-usage, month, and industry fixed effects. The controls are land quality, logarithm of land area sold, firm size, sale method, and state-ownership status and ``its interaction with transactions after 2012, central inspection, Xi-appointed officials, and pre-2012 inspection" (p. 216). Standard errors (in parentheses) are robust to two-way clustering by firm and province (\sym{*}\(p<0.05\), \sym{**}\(p<0.01\), \sym{***}\(p<0.001\)).
}
\end{table}

\noindent whether a central inspection occurred in the province at the time of sale, whether the transaction occurred after the anticorruption campaign began, etc.). As with CK's Table V, these results are robust even in the presence of potentially erroneous duplicate rows. 

\vspace{-4mm}
\section{Misspecifying Area} \label{misspecifying area}
\vspace{-2mm}
For their later tables, CK filter and transform the land transaction records to group by relevant features. When they aggregate transaction records at the firm and year level\textemdash such that they list the total area of all of firm $k$'s transactions in a given year\textemdash the value that they define as the natural logarithm of the area purchased in their paper is actually the area purchased ($\text{m}^2$) divided by 1,000,000, per their code. However, CK interpret these area/1,000,000 values as logarithms; I include specific examples in the  appendix. Thus, I re-interpret their coefficients for the tables impacted by this transformation error, for cases with and without duplicates. This error extends to CK's Tables VI, VIII, IX, XI, and XII as well as Figure VI.\footnote{The previously discussed Tables V and X do not suffer from these issues because they do not contain aggregated measures of the logarithm of area, remaining at the parcel-level. The remaining tables (I, II, III, IV, and VII) offer summary information and are not impacted by the misinterpretation; of CK's figures, only Figure VI is impacted.} 
\vspace{-5mm}
\subsection{Replicating CK's Table VI}
\vspace{-2mm}
CK's Table VI ``Quantity of Land Purchased by Princeling Firms, 2004-2016" regresses the area of land purchased in the primary land market, divided by 1,000,000, on the dummy indicator signaling a princeling connection, replicated in Panel A of Table \ref{tab: Table 6 original area/10000}.\footnote{While CK state that industry fixed effects are included in Table VI, they are omitted from their code.} While CK originally find that princeling-connected firms purchased 0.2\% ((exp(0.002)$-1)\times100)$ more land each year than their nonprinceling counterparts, the regression coefficient actually suggests that princeling-connected firms purchased an additional 2,000 ($0.002 \times $1,000,000) square meters of land annually compared to their nonprinceling peers, across all land transactions combined. Contextually, 90.8\% of rows have a total annual area purchased of zero square meters, as few firms (only 14\%) purchase land in multiple years between 2004 and 2016. Excluding firms who purchase zero area, the median area of land purchased annually by a firm is 26,668 square meters; the mean is 68,193 square meters, as a small number of firms have very high yearly transaction amounts (up to 14.67 million square meters). I thus find that the median more accurately reflects the average firm’s yearly transaction volume. When the rows with zero area are included, the median is 0 square meters, and the mean is 6,250 square meters. Excluding zeros, the average number of parcels purchased annually per firm is 1, although the maximum is 826 parcel purchases. 

The 2,000 square meter advantage is therefore a 7.5\% (2,000/26,668) increase over the median parcel size (excluding rows with zero area), much more than CK's original 0.2\%. In column (2),  firms with PSCM connections purchase 32,000 square meters ((0.001+0.031)$\times1,000,000)$ of additional land per year compared to nonprincelings\textemdash more than doubling the median parcel size; this is far above the 3\% that CK originally obtained (column (2), (exp(0.001+0.031)$-1)\times100)$). Column (3) indicates that retirement has a negligible effect on the quantity of land purchased. 

When Table VI is recalculated without duplicates (Table \ref{tab: Table 6 original area/10000}, Panel B), the result of column (1) does not change, but column (2) now suggests that princelings with PSCM connections purchase 41,000 (0.041$\times$1,000,000) additional square meters of land a year than their nonprinceling peers. Retired official connections (column (3)) are now associated with a significant increase in the area of land purchased by princelings, with princeling firms connected to retired officials purchasing, on average, an additional 59,000 ((0.001+0.058)$\times$1,000,000) square meters of land relative to nonprincelings\textemdash bringing their total purchases to more than three times the median parcel size.\footnote{Note that a relatively small number of firms (668) are connected to retired officials, and the data suggests that among purchases made by firms with connections to retired officials, smaller parcels are disproportionately represented among the duplicates, causing this large increase in the coefficient. Of the unique transactions involving retired officials, 46\% have at least one duplicate, with some having as many as 22 duplicates. Indeed, the mean and median yearly parcel purchase areas are significantly higher for firms connected to retired officials while those for nonprincelings and for princelings without this connection are roughly equal to each other.} 

\vspace{3mm}
\begin{table}[!h] 
\centering
\caption{CK's Table VI}
\label{tab: Table 6 original area/10000}
\small
\begin{threeparttable}
\setlength{\tabcolsep}{6pt} 
\renewcommand{\arraystretch}{1.3} 
\begin{tabular}{l*{6}{c}}
\hline \hline
& \multicolumn{6}{c}{Area of land purchased ($\text{m}^2$)/1,000,000} \\
\cmidrule(lr){2-7}
                    & \multicolumn{3}{c}{Panel A: Original data} & \multicolumn{3}{c}{Panel B: Data without duplicates} \\
\cmidrule(lr){2-4} \cmidrule(lr){5-7}
                    & (1) & (2) & (3) & (1) & (2) & (3) \\

\midrule
Princeling firms         & 0.002\sym{***} & 0.001\sym{**} & 0.002\sym{***} & 0.002\sym{***} & -0.000 & 0.001\sym{**} \\
                         & (0.000) & (0.000) & (0.000) & (0.000) & (0.000) & (0.000) \\
Princeling firms*PSCM &   & 0.031\sym{***} &  & & 0.041\sym{***} &  \\
                         & & (0.004)     &       & & (0.004)  \\
Princeling firms*Retired &  &  & -0.001 &  &  & 0.058\sym{***} \\
                         &     &     & (0.001) &       &       & (0.004) \\
\midrule
Adjusted \(R^{2}\)       & 0.015 & 0.016 & 0.015 & 0.009 & 0.010 & 0.010 \\
\hline \hline
\end{tabular}
\end{threeparttable}
\caption*{\footnotesize Note: Following CK, all columns include year fixed effects, as well as the control variables of firm size and state-ownership status. Standard errors are robust to clustering by firm and are in parentheses (\sym{*} \(p<0.05\), \sym{**} \(p<0.01\), \sym{***} \(p<0.001\)). There are 5,690,984 observations in each column. }
\end{table}

\vspace{-5mm}
\subsection{Tables VIII, IX, and XII}
\vspace{-2mm}
For Tables VIII, IX, and XII, I am unable to replicate CK's aggregated area measures. These specifications focus on whether there is an association between officials selling land to princelings for discounted prices and their likelihood of promotion. As such, they regress political turnover for provincial officials on different measures of princeling connections (Table VIII), political turnover for municipal officials (Table IX), and then political turnover for both types of officials (Table XII); they thus aggregate the total land sold to firms under each official's purview by year and province/municipality, respectively. CK likely has another variable that connects a transaction to an official at the province/municipality level, but they exclude this linking variable from the replication data, preventing recalculation.\footnote{\textcite{wiebe_replicating_2024} was able to replicate part of CK's Table IX, specifically the coefficient on GDP growth in columns (7) and (8); Wiebe noted that he received additional data from one of the authors.} Yet, in all cases, the logarithm of the area sold seems extremely small, suggesting that this ``logarithm of area sold" is again area ($\text{m}^2$) /1,000,000.\footnote{For Table XII, CK utilize the logarithm of area as the independent variable for a set of regressions, calling it $AreaofLandPurchased_{it}$, yet as this variable is described as measuring the ``quantity of land purchased," it is not clear whether they mean to interpret it as the logarithm of area or as area in square meters (p. 220). Regardless, as CK do not offer a detailed interpretation of these coefficients, the substantive implications do not change: the effect of $AreaofLandPurchased_{it}$ on the promotion of party secretaries drops for transactions occurring after 2012 for provincial and municipal secretaries and in regions with central inspections for provincial party secretaries.} 

\vspace{-5mm}
\subsection{Table XI and Figure VI}
\vspace{-2mm}
CK's Table XI investigates whether the quantity of land purchased by princeling firms changed after President Xi took office and began the anticorruption campaign.\footnote{One of Xi's first initiatives as party leader was the anticorruption campaign: he announced at the 18th National Congress (the top meeting for setting agendas within the party) that he would work to catch the ``tigers" (high-ranking corrupt officials) and ``flies" (low-ranking corrupt officials) to ensure the party's survival (\cite{wong_new_2012}; \cite{xi_power_2013}; \cite{yuen_disciplining_2014}).} Table XI thus regresses the land area purchased ($\text{m}^2$) /1,000,000 on whether a transaction occurred after 2012 (column (1)), whether a central inspection took place (column (2)), and whether the transaction occurs in a province wherein Xi replaced the party secretary (column (3)); these results are illustrated in Table \ref{tab:Table 11 dups area10k}. CK originally found that princeling firms purchased more land than their nonprinceling counterparts by approximately 8.1\% ((exp(0.078)$-1)\times100$), but this advantage dropped by 2.3\% ((exp($0.078)-$exp($0.078-0.022$))$\times100$) after 2012. Reinterpreting these coefficients yields that princeling firms annually purchased 78,000 ($0.078\times$1,000,000) more square meters of land than their nonprinceling counterparts, on average, before 2012, dropping to 56,000 ($(0.078-0.022)\times$1,000,000) square meters afterwards. With the median total firm purchase in a year/province being 26,114 square meters, these purchase advantages reflect a sizeable princeling benefit, suggesting that princelings purchased almost quadruple the median land purchase volume before 2012 and then only triple the median after 2012.\footnote{The panel data used in Table XI lists each firm's parcel purchases grouped by each province and year between 1991-2016. Given few firms make frequent land purchases, 95.4\% of the observations have zero square meters of land purchased in a given province/year. The reported median includes only those nonzero transactions as the median would be 0 otherwise. The mean is 3,076 square meters when including rows with zero transactions per year/province and 66,423 square meters when these zeros are excluded.}

In column (4), CK originally find that the binary variable reflecting whether a transaction occurs after 2012 becomes insignificant when the dummy variables for central inspection and Xi-appointed officials are included. They further note that ``given that the sum of the coefficients in column (4) is not significantly different from zero, these two measures together effectively eliminate the advantage of the princeling firms in purchasing a larger quantity of land" (p. 219).\footnote{Here, CK sum all coefficients including $PrincelingFirms$\textemdash that is, the dummy variable and its interactions.} This result remains in the absence of the intended logarithm.

When duplicates are removed (Table \ref{tab:Table 11 dups area10k}, Panel B), the magnitude of the princeling effect drops across all specifications but remains statistically significant: column (1), for instance, indicates that princeling firms purchased 48,000 ($0.048\times$1,000,000) square meters more land than nonprincelings annually. After 2012, this effect does not change in a statistically significant way. For column (4), as above, ``transactions after 2012" is insignificant when ``central inspection" and ``Xi-appointed officials" are included, the coefficient changing sign now that duplicates have been removed, and the sum of the princeling dummy variable and its interacted terms is not statistically different from zero. ``Transactions after 2012" is thus no longer significant in any specification, suggesting that the effect of the anticorruption campaign may not be as clear-cut.

\vspace{3mm}
\begingroup
\setlength{\tabcolsep}{3pt} 
\renewcommand{\arraystretch}{0.95}

\begin{table}[!htpb]
\centering
\caption{CK's Table XI}
\label{tab:Table 11 dups area10k}
\resizebox{\textwidth}{!}{ 
\begin{threeparttable}
\setlength{\tabcolsep}{6pt} 
\renewcommand{\arraystretch}{1.3} 
\begin{tabular}{l*{8}{c}}
\hline \hline
& \multicolumn{8}{c}{Area of land purchased ($\text{m}^2$)/1,000,000} \\
\cmidrule(lr){2-9} 
& \multicolumn{4}{c}{Panel A: Original data} & \multicolumn{4}{c}{Panel B: Data without duplicates} \\
\cmidrule(lr){2-5} \cmidrule(lr){6-9}
                    & (1) & (2) & (3) & (4) & (1) & (2) & (3) & (4) \\
\midrule
Princeling firms        & 0.078\sym{***} & 0.073\sym{***} & 0.075\sym{***} & 0.078\sym{***} & 0.048\sym{***} & 0.050\sym{***} & 0.050\sym{***} & 0.048\sym{***} \\
                         & (0.008) & (0.007) & (0.007) & (0.008) & (0.005) & (0.005) & (0.005) & (0.005) \\
Princeling firms & -0.022\sym{**} & &  & -0.015   &0.006 & & & 0.013  \\
\hspace{15pt} $*$Transactions after 2012                         & (0.006) & & & (0.007)    & (0.005) &       & & (0.007)    \\
Central inspection       & & 0.038\sym{***} & & 0.025\sym{***} & & 0.038\sym{***} & & 0.025\sym{***}  \\
                         & & (0.002) & & (0.004) &  &      (0.001) & & (0.004)      \\
Princeling firms   &  & -0.053\sym{***} & & -0.025\sym{***} &  & -0.032\sym{***} & & -0.024\sym{**}  \\
\hspace{15pt} $*$Central inspection                         & & (0.007) & & (0.007)   &       & (0.006) & & (0.007)      \\
Xi-appointed officials   & & & 0.036\sym{***} & 0.028\sym{***} & & & 0.036\sym{***} & 0.029\sym{***}  \\
                         & & & (0.002) & (0.002)     &   & &(0.002) & (0.003)   \\
Princeling firms & & & -0.056\sym{***} & -0.036\sym{***} & & & -0.033\sym{***} & -0.035\sym{***} \\
\hspace{15pt} $*$Xi-appointed officials                         &&& (0.006) & (0.006) &   & & (0.005) & (0.006)      \\
\midrule 
Adjusted \(R^{2}\)  & 0.032       & 0.047       & 0.052       & 0.057   & 0.021       & 0.038       & 0.044       & 0.050       \\
\hline \hline
\end{tabular}
\end{threeparttable}
}
\caption*{\footnotesize Note: Following CK, all columns include province and year fixed effects, as well as the control variables of firm size and state-ownership status. Standard errors are robust to two-way clustering by firm and province and are in parentheses (\sym{*} \(p < 0.05\), \sym{**} \(p < 0.01\), \sym{***} \(p < 0.001\)). All columns have 11,516,622 observations.
}
\end{table}

\vspace{-1mm}
Finally, the misinterpreted logarithm of area sold is also carried into CK's Figure VI. Originally, CK find a consistent positive difference in the quantity of land purchased between princeling and nonprinceling firms, with the 95\% confidence interval excluding 0 from 2007-2012, as is apparent in Figure \ref{fig:fig6_with_duplicates_10k}. The quantity difference then ``shrunk noticeably" after 2012, with the effect being statistically insignificant in 2013 and only marginally significant in 2015 and 2016 (p. 219). Reinterpreting the y-axis yields a smaller range of differences between princeling and nonprinceling land purchases, with the differences ranging from slightly below 0 to nearly 8,000 square meters when the outer bounds of the confidence interval are included.

In Figure VI without duplicates (illustrated in Figure \ref{fig:fig6_no_duplicates_10k}), the effect of the anticorruption campaign is less clear: the quantity difference in land purchased between princeling and nonprinceling firms drops in 2011 prior to the start of the anticorruption campaign, although the confidence intervals remain large. Further, the 95\% confidence interval of the quantity difference between princelings and nonprincelings does not drop fully towards zero as it did in Figure \ref{fig:fig6_with_duplicates_10k}, with the quantity difference remaining statistically significant after 2007. The effect is ultimately much less clean, again limiting arguments about the anticorruption campaign's causal impact.

\vspace{-6mm}
\section{Discussion}
\vspace{-2mm}
The presence of duplicates and the misinterpreted logarithm of area raise important implications about CK's original results: while the princeling effect often remained statistically significant but shrunk when duplicates were removed (as apparent in Table \ref{tab:Table 11 dups area10k}), the economic significance is not fully clear. Given the significant variation in firm parcel purchase volumes and sizes, a benefit of around 78,000 square meters (Table \ref{tab:Table 11 dups area10k}, column (1)) is extremely significant for a firm whose yearly purchase total per province is around the median of 26,114 square meters, but less so for one whose yearly purchase volume is over 4 million square meters (the year-level maximum).

Importantly, these results also raise questions about underlying data quality; with the panel data used in CK's Tables VI and XI having approximately 91\% and 95\% zeros, respectively, in the area column, effects are estimated from a small number of nonzero coefficients and are sensitive to duplicates. Given these consistency issues, taking the logarithm of the area would certainly afford more meaningful insight. Yet, when the logarithm of area is used (as described in the  appendix), the results are unstable, changing sign and direction across specifications.\footnote{In the  appendix, the logarithm is calculated as log(y+1), as this is the method that CK utilize in their paper. Given log(y+1) assigns an arbitrary weight to the extensive margin effect, I also estimate these specifications using Poisson regression and still find unstable effect estimates. I further examine the extensive margin effect via logistic regression and a linear probability model. } 

Further, the causal implications of the anticorruption campaign are limited when duplicates are removed, with Figure \ref{fig:fig6_no_duplicates_10k} highlighting that the difference in princeling and nonprinceling land purchase volumes dropped before 2012. Similarly, Table \ref{tab:Table 11 dups area10k} confirms that the coefficient for a transaction occurring after 2012 changes sign and is no longer statistically significant\textemdash although the variables representing central inspections and Xi-appointed officials remain significant\textemdash perhaps suggesting that the anticorruption campaign depends more on regional enforcement and signaling. Given these results as well as the instability of the analysis when the logarithm is used (per the  appendix), further investigation should be done to determine whether the anticorruption campaign truly plays a causal role in reducing princeling discounts. 

\vspace{4mm}
\begin{figure}[H]
    \centering
    \caption{Comparison of CK's Figure VI with and without duplicates}
    \begin{subfigure}[t]{0.47\linewidth}
        \centering
        \includegraphics[width=\linewidth]{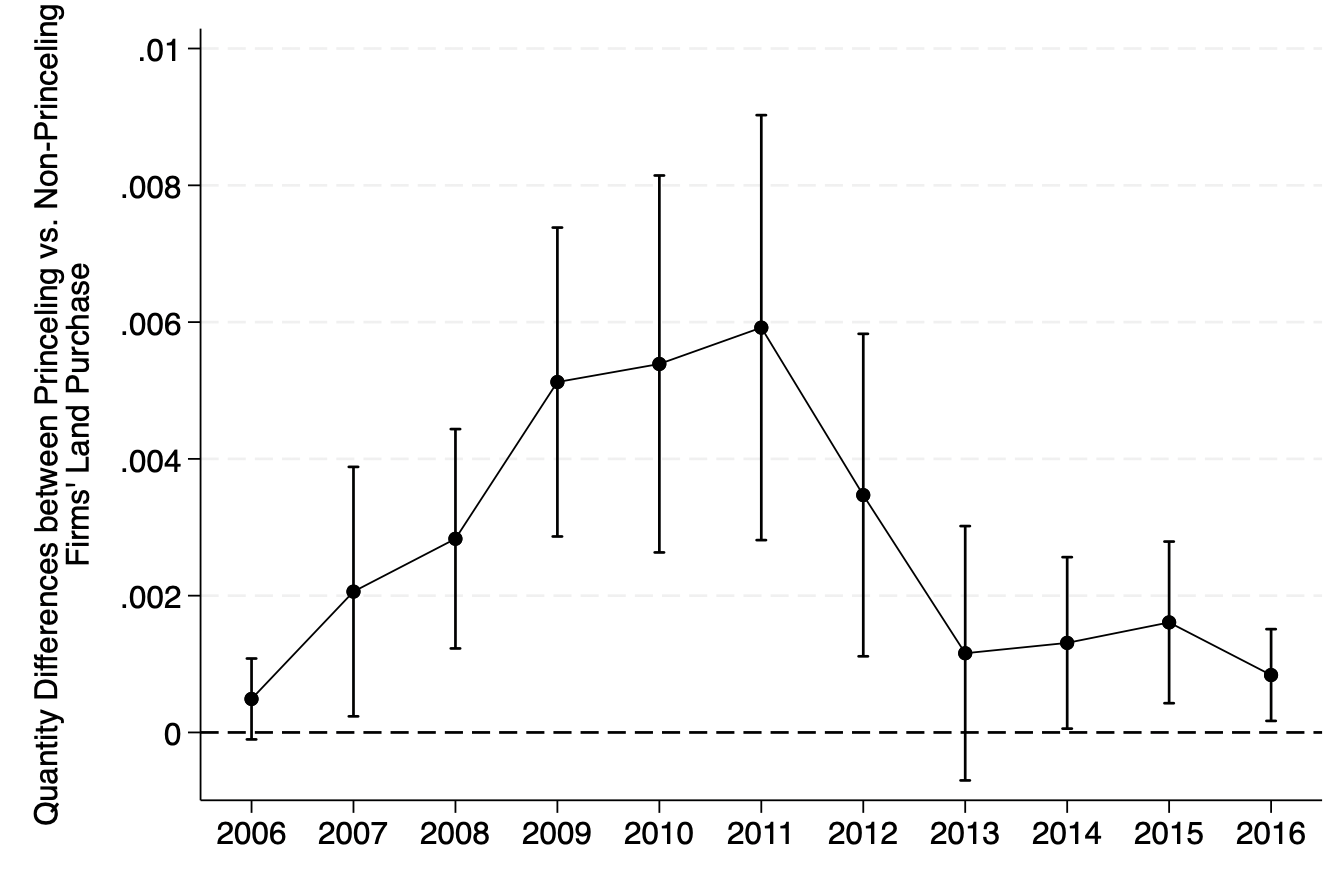}
        \caption{CK's Figure VI}
    \label{fig:fig6_with_duplicates_10k}
    \end{subfigure}%
    \hspace{0.05\linewidth} 
    \begin{subfigure}[t]{.47\linewidth}
        \centering
        \includegraphics[width=\linewidth]{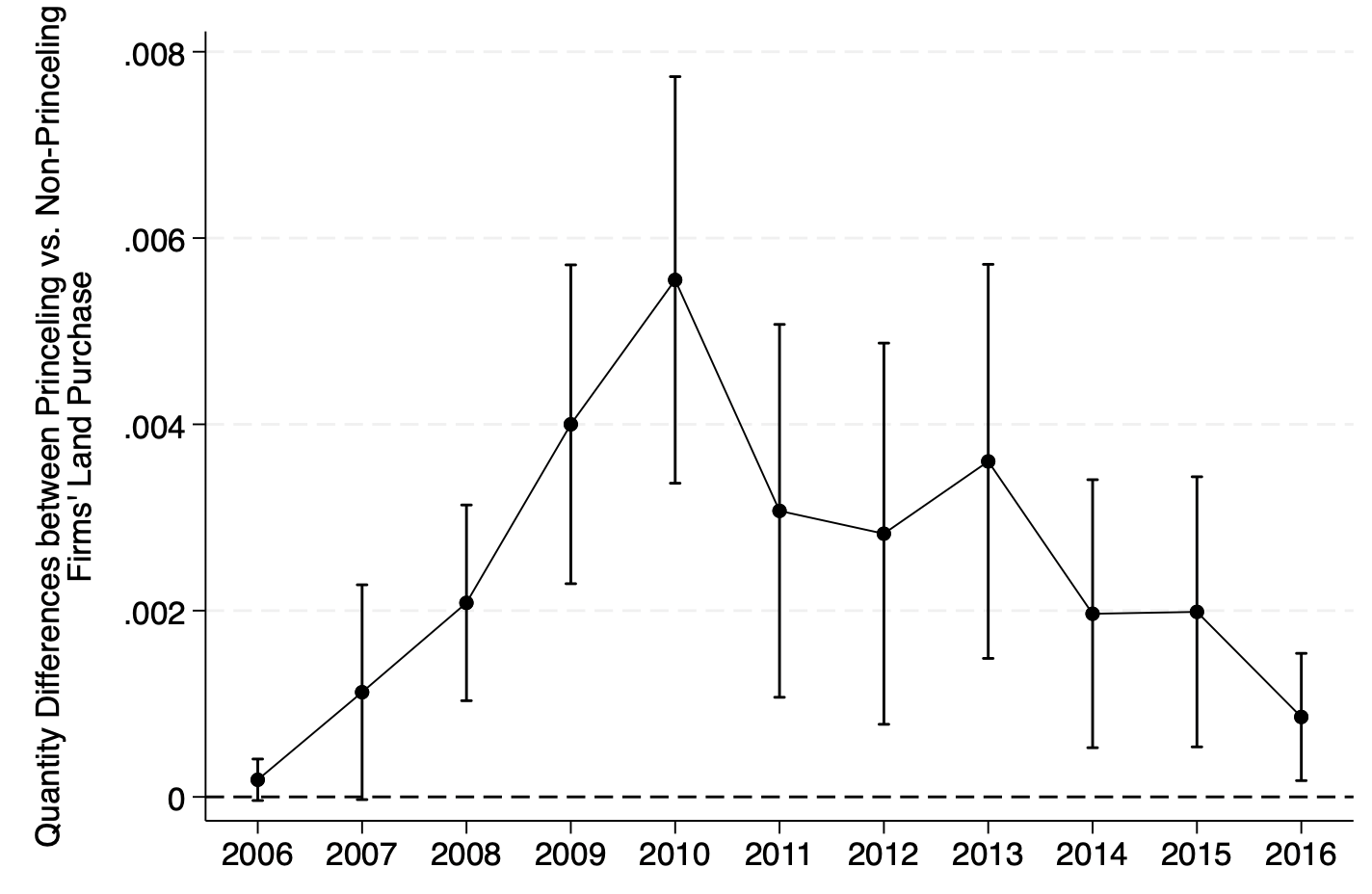}
        \caption{CK's Figure VI, no duplicates}
        \label{fig:fig6_no_duplicates_10k}
    \end{subfigure}
    \caption*{\footnotesize \raggedright Note: Bars reflect 95\% confidence intervals.}
    \label{fig:comparison_fig6_2}

\end{figure}

\pagebreak

\printbibliography

\pagebreak
\appendix
\section*{Appendix}
\global\long\def\thesection{A}%
\global\long\def\theequation{A.\arabic{equation}}%
\global\long\def\thetable{A\arabic{table}}%
\global\long\def\thefigure{A\arabic{figure}}%
\setcounter{table}{0}\setcounter{equation}{0}\setcounter{figure}{0}
\subsection{Duplicate examples}

Example of correct duplicates (two parcels that are right next to each other, having the same buyer, transaction month/year, and parcel area): princeling transaction IDs 3060341 and 3060345, which have electronic lookup numbers 6543222013B00600 and 6543222013B00595. The Transfer Announcements confirm that they are different parcels.
\medskip
\newline Incorrect duplicates:
\begin{enumerate}
    \item Duplicate listings with multiple parcels are assigned the same electronic lookup number, with the Transfer Announcement confirming there is only one parcel of the given size for sale. 
    \newline Example: Princeling transaction IDs: 125957, 1527560. \newline Electronic lookup number (for both listings in the database): 2106242010B00016.
    \item Electronic lookup numbers for multiple parcels are the same, although Transfer Announcements are not available to confirm whether the parcels are indeed the same. \newline Example: Princeling transaction IDs: 135826, 2991138. \newline Electronic lookup number (for both listings in the database): 2112232010B00057. 
    \item Duplicate listings have multiple electronic lookup numbers, but the Transfer Announcement suggests they refer to the same underlying parcel. In this case, one listing appears to be more complete than the other.
    \newline Example: Princeling transaction IDs: 3002177, 105512, 943888. \newline Electronic lookup numbers: 1525302010B00059, 1525302010B00076.
    \newline Note that in this case, there are three princeling transactions with the parcel features but only two listings in the Ministry of Land's database, suggesting there are two erroneous duplicates. 
    \item Duplicate listings have high numbers of duplicates that are nonsensical and are sometimes wholly omitted from the database. 
    \newline Example 1: 73 entries of a parcel with an area of 100 square meters purchased by the same buyer at the same price in the same month/year/province (all princelings). 
    \newline Princeling transaction IDs from these listings include: 508039, 508046, 508048, 508049, 508063, ....
    \newline Example 2: 71 duplicates with an area of 289 square meters.  
    \newline Princeling land transaction IDs: 321688, 321689, 321692, 321693, 321694, 321697, ....
    \item Partially correct duplicates exist, wherein there are two parcels right next to each other, having the same buyer, transaction month/year, and area of the parcels. The Transfer Announcements suggest they are different parcels, but there are extra copies in the princeling database. 
    \newline Example: Princeling transaction IDs: 2248276, 779393, and 1368322. \newline Electronic lookup numbers: 5301262010B00026, 5301262010B00019. 
\end{enumerate}

\subsection{Misspecifying area/1,000,000 as log(area) } \label{appendix: misspecifying area}

In CK's replication data, the dataset \texttt{price.dta} contains the parcel-level transaction records described in Section \ref{Apparent Duplicates in the Data}. By matching transaction records from the Land Transaction Monitoring System with CK's dataset, I am able to discern the original areas of each parcel.

In their subsequent tables, CK aggregate this parcel-level data, summing the given purchases by each firm in each year (Table VI), by each firm in each year in each province (Table XI), and then by each year and province/municipality (Tables VIII, IX, and XII). At this step, the misinterpretation of area becomes apparent in a simple example about firm 1.

In the year 2015, firm 1 has only one transaction with a raw area of 17,795.3 square meters. Thus, when CK aggregate the total amount of area sold to each firm in each year, firm 1's total for 2015 is still just 17,795.3 square meters. Yet, in their dataset, the value that CK has for the reported $lnarea$ is 0.01764, which is approximately 17,795.3/1,000,000. The slight variation in numbers seems to come from their calculation: they likely take the logarithm of the original area divided by 100. They then truncate/round it, exponentiate it, and divide it by 10,000.

\vspace{3mm}
\begin{table}[h!]
\centering
\caption{Sample data frame row 1}
\label{tab:Appendix table firm 1}
\resizebox{\textwidth}{!}{ 
\begin{tabular}{@{}>{\centering\arraybackslash}p{2cm} 
                >{\centering\arraybackslash}p{2cm} 
                >{\centering\arraybackslash}p{3cm} 
                >{\centering\arraybackslash}p{3cm} 
                >{\centering\arraybackslash}p{3cm} 
                >{\centering\arraybackslash}p{3cm}@{}} 
\hline \hline
Firm ID & {Year} & {Area} & {Area/1,000,000} & {CK's reported \textit{lnarea}} & {Actual \textit{lnarea}} \\
\midrule
1                & 2015          & 17,795.3          & 0.017795                    & 0.01764                      & 9.78669                \\
\hline \hline
\end{tabular}
} 
\caption*{}
\end{table}

A second example with aggregation confirms that this is the case. For instance, firm 83 has 2 land transactions that occur in 2013. I include both of these transaction records, as well as the aggregated entry, in the two panels of Table \ref{tab:Appendix table firm 83}.

\begin{table}[H]
\centering
\caption{Raw and aggregated data for firm 83}
\label{tab:Appendix table firm 83}
\resizebox{\textwidth}{!}{ 
\begin{tabular}{@{}llcccc@{}}
\hline \hline
\multicolumn{6}{c}{\small {Panel A: Transaction-level data}} \\ 
\midrule
\small {Firm ID} & \small {Year} & \small {CK transaction identifier} & \small {Electronic lookup no.} & \small {Area} & \small {CK's reported \textit{lnarea}} \\
\midrule
83               & 2013          & 2086875                             & 1310822013B00732                          & 17481.17              & 5.17 \\
83               & 2013          & 2090254                             & 1310822013B00680                          & 639.87              & 1.86 \\
\midrule
\multicolumn{6}{c}{\small {Panel B: Data aggregated by firm}} \\ 
\midrule
\small {Firm ID} & \small {Year}  & \small {Area (Sum)} & \small {Area/1,000,000} & \small {CK's reported \textit{lnarea}} &\small {Actual \textit{lnarea}}  \\
\midrule
83               & 2013    & 18121.04       &     0.01812     & 0.01796        & 9.392     \\
\hline \hline
\end{tabular}
} 
\caption*{\footnotesize Note: The electronic lookup number and raw area are obtained directly from the Land Transaction Monitoring System. Note that as described above, CK's reported $lnarea$ is calculated as log(area/100), and this is reflected in their reported $lnarea$ measures in Panel A. It is these values that they then round, exponentiate, and scale as they aggregate to obtain the slight differences between their reported $lnarea$ value and the area/1,000,000. Also note that the difference is slightly more pronounced in cases of firms having purchased multiple parcels in a given year, as the rounding/truncation is conducted on each individual parcel.}
\end{table}

In these cases and all others, CK's reported $lnarea$ is approximately area/1,000,000, the slight difference again resulting from CK's rounding during intermediate steps. In Section \ref{misspecifying area}, I reinterpret CK's main tables in light of the fact that area is scaled rather than logged; the tables recalculated with correct log values are included in Appendix \ref{appendix: log(y+1)}.

\subsection{Estimating semi-elasticities} \label{appendix: log(y+1)}

In this section, I estimate what Chen and Kung claim to have done in their paper\textemdash that is, conduct the calculations with the logarithm of area purchased\textemdash and find, in many cases, substantially different results, with and without duplicates. In particular, the princeling effect changes its sign and direction, differing in cases with and without duplicates; in multiple instances, the results are qualitatively different when the logarithm of area value is used, even when calculated on the original data. Further, there is evidence of a sharp drop in the princeling advantage prior to the anticorruption campaign. As above, I regenerate CK's Tables VI and XI as well as Figure VI, which are impacted by these errors. For this analysis, CK seek to analyze the differences in quantity purchased between princelings and nonprincelings overall, not just those firms who buy land in a given year\textemdash and presumably, princelings may be more likely to purchase land in a given year in the first place. Thus, I follow CK in calculating log(area + 1), not using a scaling factor and adding 1 to the area value to ``deal with observations of 0" prior to taking the log (p. 212). However, because this approach assigns an arbitrary weight to the extensive margin effect, I also recalculate these tables using Poisson regression in Section \ref{sec: poisson regression}. I then evaluate the extensive margin effect in Section \ref{appendix: extensive margin}. 

\sloppy For Table VI, CK originally interpret the regression coefficient as suggesting that princeling-connected firms purchased 0.2\% (column (1), (exp(0.002$)-1)\times100$) more land each year than  

\begin{table}[H] 
\centering
\caption{CK's Table VI}
\label{tab: Table VI Appendix}
\small
\begin{threeparttable}
\setlength{\tabcolsep}{6pt} 
\renewcommand{\arraystretch}{1.3} 
\begin{tabular}{l*{6}{c}}
\hline \hline
& \multicolumn{6}{c}{Log of area of land purchased} \\
\cmidrule(lr){2-7}
                    & \multicolumn{3}{c}{Panel A: Original data} & \multicolumn{3}{c}{Panel B: Data without duplicates} \\
\cmidrule(lr){2-4} \cmidrule(lr){5-7}
                    & (1) & (2) & (3) & (1) & (2) & (3) \\

\midrule
Princeling firms    &       0.117\sym{***}&      -0.304\sym{***}&      -0.063\sym{***}&       0.096\sym{***}&      -0.296\sym{***}&      -0.076\sym{***}\\
                    &     (0.013)         &     (0.011)         &     (0.012)         &     (0.013)         &     (0.011)         &     (0.011)         \\
Princeling firms*PSCM&                     &       7.795\sym{***}&                     &                     &       7.241\sym{***}&                     \\
                    &                     &     (0.082)         &                     &                     &     (0.081)         &                     \\
Princeling firms*Retired&                     &                     &       8.208\sym{***}&                     &                     &       7.834\sym{***}\\
                    &                     &                     &     (0.119)         &                     &                     &     (0.115)         \\

\midrule
Adjusted \(R^{2}\)  &       0.044         &       0.046         &       0.045         &       0.044         &       0.046         &       0.045         \\
\hline \hline
\end{tabular}
\end{threeparttable}
\caption*{\footnotesize Note: Following CK, all columns include year fixed effects, as well as the control variables of firm size and state-ownership status. CK note that industry fixed effects are also included, but their code confirms they are omitted. I follow their code and do not include industry fixed effects. Standard errors are robust to clustering by firm and are in parentheses (\sym{*} \(p<0.05\), \sym{**} \(p<0.01\), \sym{***} \(p<0.001\)). There are 5,690,984 observations in each column.}
\end{table}

\noindent their nonprinceling counterparts; they find that the quantity of land purchased by firms connected to Politburo Standing Committee members (PSCM) is 3\% higher (column (2), (exp(0.001+0.031)$-1) \times100$) than that purchased by nonprincelings and that retirement has a negligible effect on the quantity of land purchased by princeling firms (column (3)). Yet, as highlighted in Panels A and B of Table \ref{tab: Table VI Appendix}, these results differ significantly from calculations that involve the actual logarithm of area values. In the baseline specification, princeling-connected firms purchase 12.4\% ((exp($0.117)-1)\times100$) more land each year than their nonprinceling counterparts. The other specifications are more complicated: column (2) suggests that princeling firms purchase 26.2\% less land ((exp($-0.304)-1)\times100$) than their nonprinceling counterparts but that princeling firms with PSCM connections purchase a massive 179,084\% more land ((exp($-$0.304+7.795)$-$1)$\times100$) than nonprinceling firms. Similarly, column (3) indicates that princelings purchase 6.1\% less land ((exp($-0.063)-1)\times100$) than their nonprinceling counterparts but that princeling firms connected to retired officials purchase 344,511\% more land ((exp($-$0.063+8.208)$-$1$)\times100$) than their nonprinceling peers. Such percentages are so large as to be nonsensical: with the median area of land purchased annually by a firm being 26,668 square meters, a 344,511\% increase above this would be 91,900,739 square meters (26,668$\times $exp($-0.063+8.208)$)) while a 179,084\% increase would be 47,784,870 square meters (26,668$\times$exp($-$0.304+7.795))\textemdash and these are well above the maximum annual firm purchase volume of around 14.7 million square meters.\footnote{As above, this median statistic includes only the rows in the panel that have nonzero area values (pre-log); otherwise, the median would be 0. The mean is 6,250 square meters with zeros and 68,193 square meters excluding zeros, as a small number of firms have very high transaction amounts and skew the mean upwards. I thus find the median more accurately reflects the average firm's yearly transaction volume but note that the princeling advantage would yield even more significant increases in area if the mean were used instead.} These specifications are also interesting because columns (2) and (3) suggest that regular princelings purchase less land than nonprincelings, but princelings with PSCM and retired official connections purchase significantly more, reversing CK's original conclusions\textemdash and the regressions calculated without duplicates (Panel B, columns (1)-(3)) further echo these results.

CK's Table XI investigates whether the quantity of land purchased by princeling firms changed after President Xi took office and began the anticorruption campaign. Table XI thus regresses the logarithm of land area purchased on whether a transaction occurred after 2012 (column (1)), whether a central inspection took place (column (2)), and whether the transaction occurred in a province wherein Xi replaced the party secretary (column (3)). CK originally found that princeling firms purchased more land than their nonprinceling counterparts by approximately 8.1\%; after 2012, princeling land purchases dropped by more than 2.3\%. When using log(area+1), as shown in Panel A of Table \ref{tab: Table 11 Appendix}, the regressions suggest princeling firms purchased 244,694\% ((exp(7.803)$-$1)$\times100$) more land than their nonprinceling counterparts prior to 2012, and after 2012, their land purchases increase by 52,112 percentage points ((exp(7.803+0.193)$-$exp(7.803))$\times100$).

\vspace{3mm}
\begin{table}[H]
\centering
\caption{CK's Table XI}
\label{tab: Table 11 Appendix}
\resizebox{\textwidth}{!}{ 
\begin{threeparttable}
\setlength{\tabcolsep}{6pt} 
\renewcommand{\arraystretch}{1.3} 
\begin{tabular}{l*{8}{c}}
\hline \hline
& \multicolumn{8}{c}{Log of area of land purchased} \\
\cmidrule(lr){2-9} 
& \multicolumn{4}{c}{Panel A: Original data} & \multicolumn{4}{c}{Panel B: Data without duplicates} \\
\cmidrule(lr){2-5} \cmidrule(lr){6-9}
                    & (1) & (2) & (3) & (4) & (1) & (2) & (3) & (4)  \\
\midrule
Princeling firms    &       7.803\sym{***}&       7.938\sym{***}&       7.905\sym{***}&       7.800\sym{***}&       7.241\sym{***}&       7.484\sym{***}&       7.414\sym{***}&       7.238\sym{***}\\
                    &     (0.205)         &     (0.189)         &     (0.205)         &     (0.207)         &     (0.205)         &     (0.193)         &     (0.211)         &     (0.207)         \\
Princeling firms&       0.193         &                     &                     &       0.770\sym{***}&       0.650\sym{***}&                     &                     &       1.182\sym{***}\\
\hspace{15pt}$*$Transactions after 2012                    &     (0.150)         &                     &                     &     (0.154)         &     (0.140)         &                     &                     &     (0.152)         \\
Central inspection  &                     &       9.014\sym{***}&                     &       5.523\sym{***}&                     &       8.958\sym{***}&                     &       5.479\sym{***}\\
                    &                     &     (0.071)         &                     &     (0.895)         &                     &     (0.081)         &                     &     (0.890)         \\
Princeling firms&                     &      -8.468\sym{***}&                     &      -5.495\sym{***}&                     &      -8.060\sym{***}&                     &      -5.442\sym{***}\\
\hspace{15pt}$*$Central inspection                    &                     &     (0.199)         &                     &     (0.925)         &                     &     (0.197)         &                     &     (0.920)         \\

Xi-appointed officials&                     &                     &       9.079\sym{***}&       7.307\sym{***}&                     &                     &       9.038\sym{***}&       7.280\sym{***}\\
                    &                     &                     &     (0.115)         &     (0.399)         &                     &                     &     (0.114)         &     (0.395)         \\
Princeling firms&                     &                     &      -8.260\sym{***}&      -7.040\sym{***}&                     &                     &      -7.806\sym{***}&      -6.946\sym{***}\\
\hspace{15pt}$*$Xi-appointed officials                    &                     &                     &     (0.205)         &     (0.439)         &                     &                     &     (0.207)         &     (0.434)         \\
                    \midrule
Adjusted \(R^{2}\)  &       0.094         &       0.198         &       0.250         &       0.283         &       0.094         &       0.203         &       0.259         &       0.294         \\
\hline \hline
\end{tabular}
\end{threeparttable}
}
\caption*{\footnotesize Note: Following CK, all columns include province and year fixed effects, as well as the control variables of firm size and state-ownership status. Standard errors are robust to two-way clustering by firm and province and are in parentheses (\sym{*} \(p < 0.05\), \sym{**} \(p < 0.01\), \sym{***} \(p < 0.001\)). All columns have 11,516,622 observations.
}
\end{table}

For the interacted central inspection equation (column (2)), the estimation suggests that the land sold to princeling firms increased by 203,501 percentage points ((exp(7.938+9.014$-$8.468)$-$exp(7.938))$\times 100)$. These percentages are so large as to be wholly nonsensical: with the median total purchased by a firm in a given year and province being 26,114 square meters, these numbers suggest the average princeling advantage would be 63,925,369 square meters (per column (1)'s 244,694\% growth, as 26,114 $\times$ exp(7.803$)=63,925,369)$\textemdash a value far above the maximum annual land purchase by a firm.\footnote{As noted in the main text, many firms purchase land infrequently, and 95.4\% of the observations have zero square meters purchased in a given province/year. The reported median includes only those nonzero transactions (as the median would otherwise be 0). The mean is 3,076 square meters when including rows with zero square meters sold and 66,423 square meters when excluding zeros.} Likewise, CK's original analysis finds that the binary variable reflecting whether a transaction occurs after 2012 becomes insignificant in column (4) when the dummy variables for central inspection and Xi-appointed officials are included, but when recalculated, this is no longer the case.

When duplicate rows are removed, as highlighted in Panel B of Table \ref{tab: Table 11 Appendix}, the coefficients are still quite large: Panel B's column (1), for instance, indicates that princeling firms purchased more land than their nonprinceling counterparts by 139,449\% ((exp(7.241)$-$1)$\times100$). After 2012, their land purchases increase by 127,763 percentage points ((exp(7.241+0.650)$-$exp(7.241))$\times100$).  Yet, due to the extremely large percentages of princeling advantage, the results are difficult to interpret and nonsensical, likely reflecting high variability in parcel sizes, the need for further controls, the relatively small number of princeling firms in the overall data, and the sensitivity resulting from adding 1 to area prior to logging it. These results are also interesting because they run counter to that from the recalculated version of CK's Table VI (Table \ref{tab: Table VI Appendix}), which suggested that princeling firms\textemdash in the presence of interacted terms\textemdash purchased less land than their nonprinceling counterparts, confirming these data and estimation issues.

In their original results for Figure VI, CK find that the consistent positive difference in the quantity of land purchased between princeling and nonprinceling firms ``shrunk noticeably" after 2012, with the effect being statistically insignificant in 2013 and only being ``marginally significant in 2016'' (p. 219). Yet, when generating CK's Figure VI with the log(area + 1) in Figure \ref{fig:fig6_with_duplicates_original}, the quantity difference between princeling and nonprinceling land purchases drops toward zero in 2011, prior to the anticorruption campaign. By 2012, this difference is no longer positive or statistically significant. Likewise, the quantity difference and its 95\% confidence interval have dropped below 0 in 2014 and 2015, suggesting princelings purchase less land than their nonprinceling counterparts, and while the drop eventually ticks back to zero in 2016, it seems that the declining trend began well before the anticorruption campaign\textemdash the effect is thus much less clean, limiting arguments about the causal impact of the anticorruption campaign. The plot without duplicates (Figure \ref{fig:fig6_no_duplicates_original}) closely echoes this behavior, again suggesting that the quantity difference between princeling and nonprinceling land purchases dropped prior to the anticorruption campaign and remained at or slightly near zero for the bulk of its duration. 

\vspace{5mm}
\begin{figure}[H]
    \centering
    \caption{Recalculating CK's Figure VI, with and without duplicates, using log(area + 1)}
    \begin{subfigure}[t]{0.45\linewidth}
        \centering
        \includegraphics[width=\linewidth]{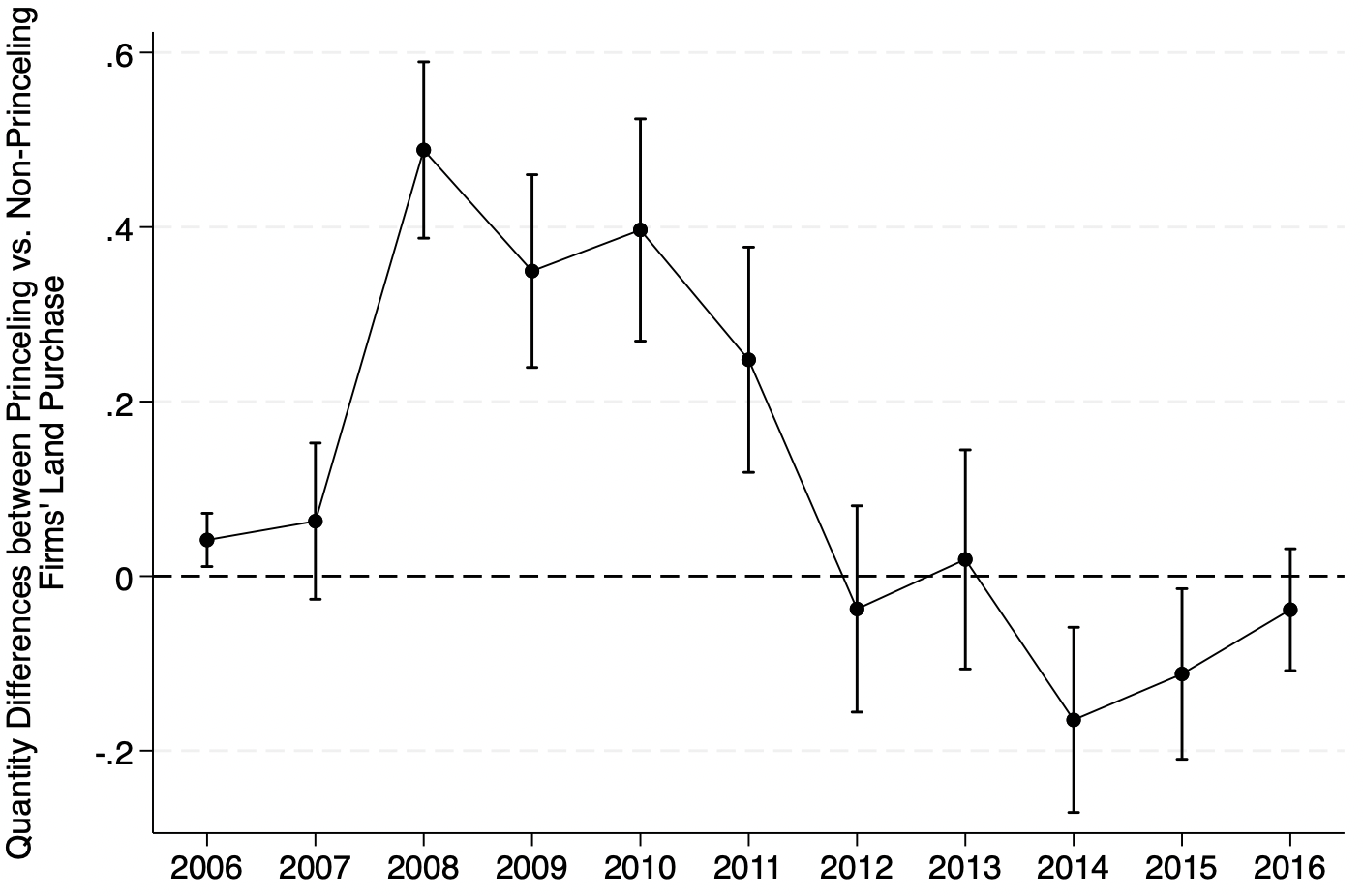}
        \caption{CK's Figure VI recalculated}
        \label{fig:fig6_with_duplicates_original}
    \end{subfigure}%
    \hspace{0.05\linewidth} 
    \begin{subfigure}[t]{0.45\linewidth}
        \centering
        \includegraphics[width=\linewidth]{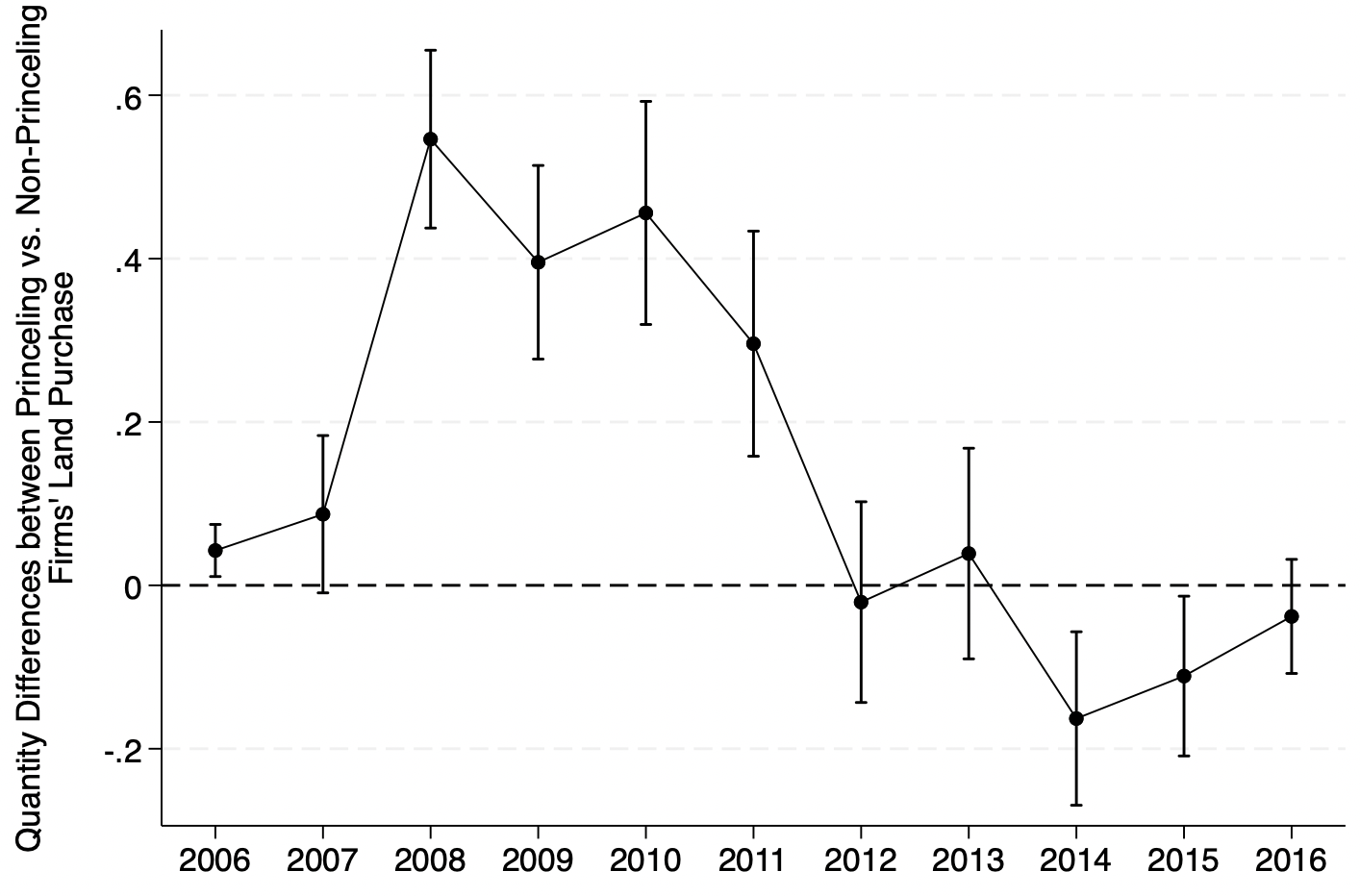}
        \caption{CK's Figure VI recalculated, no duplicates}
        \label{fig:fig6_no_duplicates_original}
    \end{subfigure}
    \caption*{\raggedright \footnotesize Note: Bars reflect 95\% confidence intervals.}
    \label{fig:comparison_fig6}
\end{figure}

\subsubsection{Poisson regression} \label{sec: poisson regression}

\textcite{chen_logs_2024} posit that if the treatment affects the extensive margin, such that $\text{P(Y(1)} = 0) \neq \text{P(Y(0) = 0)}$, the percentage treatment effect yielded by the log is not well-defined: a percentage is a unit-invariant quantity, but  log-like transformations such as log(y+1) depend on the units of the outcome. One alternative approach to determining average treatment effects in these situations is estimating scale-invariant normalized parameters, such as with Poisson regression. Via Poisson, the parameter $\theta_{ATE\%}$\textemdash which represents the ATE ``in levels expressed as a percentage of the control mean" (that is, $\theta_{ATE\%} = \frac{E[Y(1)-Y(0)]}{E[Y(0)]}$)\textemdash can be consistently estimated, yielding percentages (\cite{chen_logs_2024}, p. 913). I thus recalculate CK's Tables VI and XI, as well as Figure VI, using Poisson regression. Afterwards, I briefly evaluate the extensive margin effect in these tables and figure, using a binary variable for any nonzero purchase area as the outcome. Using \texttt{ppmlhdfe} in Stata, column (1) of Table \ref{tab: Table VI Appendix poisson} estimates 
\begin{align*}
    Y_i = \text{exp(}\beta_0 + \beta_1 PrincelingFirms_i + T_{ysm(i)}) U_i,
\end{align*}
where $T$ captures year, state-ownership status, and firm size fixed effects, and $U_i$ is a multiplicative error term with $\mathbb{E}[U_i | \cdot] = 1$. $\theta_{ATE\%}$ can then be calculated as $\theta_{ATE\%} = \text{exp}(\hat{\beta}_1) -1$. The subsequent columns of Tables \ref{tab: Table VI Appendix poisson} and \ref{tab: Table 11 Appendix poisson} follow this structure, with the appropriate interaction terms included as necessary.

\vspace{3mm}
\begin{table}[H] 
\centering
\caption{CK's Table VI, Poisson}
\label{tab: Table VI Appendix poisson}
\small
\begin{threeparttable}
\setlength{\tabcolsep}{6pt} 
\renewcommand{\arraystretch}{1.3} 
\begin{tabular}{l*{6}{c}}
            \hline \hline
& \multicolumn{6}{c}{Area of land purchased ($\text{m}^2$) } \\
\cmidrule(lr){2-7}
                    & \multicolumn{3}{c}{Panel A: Original data} & \multicolumn{3}{c}{Panel B: Data without duplicates} \\
\cmidrule(lr){2-4} \cmidrule(lr){5-7}
                    & (1) & (2) & (3) & (1) & (2) & (3) \\

\midrule
Princeling firms  &       0.540\sym{***}&       0.045         &       0.297\sym{***}&       0.399\sym{***}&      -0.035         &       0.177\sym{***}\\
            &     (0.044)         &     (0.047)         &     (0.051)         &     (0.045)         &     (0.045)         &     (0.052)         \\

Princeling firms*PSCM        &                     &       2.165\sym{***}&                     &                     &       2.072\sym{***}&                     \\
            &                     &     (0.088)         &                     &                     &     (0.093)         &                     \\
[1em]
Princeling firms*Retired     &                     &                     &       2.272\sym{***}&                     &                     &       2.142\sym{***}\\
            &                     &                     &     (0.080)         &                     &                     &     (0.080)         \\
\hline
Pseudo \(R^{2}\)  &       0.113         &       0.115         &       0.114         &       0.103         &       0.105         &       0.104         \\

\hline\hline
\end{tabular}
\end{threeparttable}
\caption*{\footnotesize Note: Following CK, all columns include year fixed effects, as well as the control variables of firm size and state-ownership status. CK note that industry fixed effects are also included, but their code confirms that they are omitted. I follow their code and do not include industry fixed effects. Standard errors are robust to clustering by firm and are in parentheses (\sym{*} \(p<0.05\), \sym{**} \(p<0.01\), \sym{***} \(p<0.001\)). There are 5,690,984 observations in each column.}
\end{table}

The basic princeling effect of column (1) becomes a 71.6\% increase over nonprincelings (($\text{exp}(0.540) -1)\times100$) when duplicates are included and a 49.0\% ($(\text{exp}(0.399) -1)\times100)$ increase over nonprincelings excluding duplicates, compared to the 0.2\% which CK originally reported in their paper. Column (2) reports a non-statistically significant coefficient on princeling firms alone, but that princeling firms with PSCM connections purchase on average 811.6\% more land (($\text{exp}(0.045+2.165)-1 )\times100$) than nonprincelings, when including duplicates. When duplicates are excluded, this becomes a 666.8\% (($\text{exp}(-0.035+2.072)-1)\times 100)$ increase over nonprincelings, with the coefficient on princeling firms without PSCM connections being negative but statistically insignificant. Column (3) indicates that princelings purchase 34.6\% (($\text{exp}(0.297)-1)\times100)$  more land than their nonprinceling counterparts, but that princeling firms connected to retired officials purchase 1,205.3\% (($\text{exp}(0.297+2.272)-1)\times 100)$ more land than their nonprinceling peers. When duplicates are excluded, these numbers become 19.4\% (($\text{exp}(0.177)-1) \times100)$ and 916.6\% (($\text{exp}(0.177 + 2.142)-1)\times100)$, respectively. While using Poisson yields no statistically significant negative coefficients on the princeling firms, as was the case when using log(y+1), the percentage increases are still so large as to be concerning: with the median area of land purchased annually by a firm being 26,668 square meters, a 666.8\% increase (per column (2), without duplicates) would be 204,478.8 square meters (26,668$\times \text{exp}(-0.035+2.072)$). While this number is now below the maximum annual firm purchase volume of 14.7 million square meters and is theoretically possible, it is still suspiciously large given the median. 

\vspace{5mm}
\begin{table}[H]
\centering
\caption{CK's Table XI, Poisson}
\label{tab: Table 11 Appendix poisson}
\resizebox{\textwidth}{!}{ 
\begin{threeparttable}
\setlength{\tabcolsep}{6pt} 
\renewcommand{\arraystretch}{1.3} 
\begin{tabular}{l*{8}{c}}
\hline \hline
& \multicolumn{8}{c}{Area of land purchased ($\text{m}^2$)} \\
\cmidrule(lr){2-9} 
& \multicolumn{4}{c}{Panel A: Original data} & \multicolumn{4}{c}{Panel B: Data without duplicates} \\
\cmidrule(lr){2-5} \cmidrule(lr){6-9}
                    & (1) & (2) & (3) & (4) & (1) & (2) & (3) & (4)  \\
Princeling firms    &       2.328\sym{***}&       2.377\sym{***}&       2.384\sym{***}&       2.328\sym{***}&       2.144\sym{***}&       2.251\sym{***}&       2.247\sym{***}&       2.144\sym{***}\\
                    &     (0.100)         &     (0.094)         &     (0.096)         &     (0.102)         &     (0.103)         &     (0.096)         &     (0.097)         &     (0.107)         \\
Princeling firms &       0.027         &                     &                     &       0.541\sym{***}&       0.202\sym{*}  &                     &                     &       0.671\sym{***}\\
\hspace{15pt}$*$Transactions after 2012                     &     (0.097)         &                     &                     &     (0.146)         &     (0.097)         &                     &                     &     (0.140)         \\
Central inspection  &                     &       2.571\sym{***}&                     &       1.367\sym{***}&                     &       2.607\sym{***}&                     &       1.341\sym{***}\\
                    &                     &     (0.058)         &                     &     (0.361)         &                     &     (0.066)         &                     &     (0.367)         \\
Princeling firms &                     &      -2.010\sym{***}&                     &      -1.165\sym{***}&                     &      -1.900\sym{***}&                     &      -1.107\sym{***}\\
\hspace{15pt}$*$Central inspection                     &                     &     (0.131)         &                     &     (0.319)         &                     &     (0.124)         &                     &     (0.314)         \\
Xi-appointed officials &                     &                     &       2.782\sym{***}&       2.233\sym{***}&                     &                     &       2.878\sym{***}&       2.342\sym{***}\\
                    &                     &                     &     (0.158)         &     (0.312)         &                     &                     &     (0.157)         &     (0.314)         \\
Princeling firms &                     &                     &      -2.053\sym{***}&      -1.977\sym{***}&                     &                     &      -1.929\sym{***}&      -1.967\sym{***}\\
\hspace{15pt}$*$Xi-appointed officials                     &                     &                     &     (0.094)         &     (0.299)         &                     &                     &     (0.094)         &     (0.293)         \\
                    \midrule
Pseudo \(R^{2}\)  &          0.123      &       0.160          &     0.179          &       0.188         &       0.113         &       0.164         &        0.192         &        0.204         \\
\hline \hline
\end{tabular}
\end{threeparttable}
}
\caption*{\footnotesize Note: Following CK, all columns include province and year fixed effects, as well as the control variables of firm size and state-ownership status. Standard errors are robust to two-way clustering by firm and province and are in parentheses (\sym{*} \(p < 0.05\), \sym{**} \(p < 0.01\), \sym{***} \(p < 0.001\)). All columns have 5,758,311 observations.
}
\end{table}

Here, Table \ref{tab: Table 11 Appendix poisson}, Panel A column (1) suggests that princeling firms purchased 925.7\% (($\text{exp}(2.328)-1)\times 100 $) more land than their nonprinceling counterparts prior to 2012, and there is no statistically significant change after 2012. For the interacted central inspection equation (column (2)), the estimation suggests that the land sold to princeling firms increased by 1787.8\% $((\text{exp}(2.377+2.571-2.010)-1)\times 100 $) in the presence of central inspections, compared to nonprincelings. With the median total area purchased by a firm in a given year and province being 26,114 square meters, this equates to a princeling advantage of 492,981.5 square meters (26,114$\times\text{exp}(2.377+2.571-2.010)$)\textemdash an increase so large as to be suspect. Columns (3) and (4) likewise yield extremely large percentage increases for princelings; in particular, when all three measures of the anticorruption campaign are included in column (4), the princeling effect remains significant and positive despite the anticorruption campaign\textemdash the opposite of CK's original conclusion.

In Panel B, which excludes duplicates, column (1) now experiences a statistically significant increase in the princeling effect after 2012: prior to 2012, princelings had a 753.4\% ((exp$(2.144)-1) \times 100$) advantage over nonprincelings, but after 2012, this increased to 944.4\% ((exp$(2.144+0.202)-1)\times 100$)\textemdash the opposite direction of CK's original finding. Panel B's column (2) suggests that princelings, in provinces with central inspections at the time of sale, purchased on average 1,825.9\% more land ((exp$(2.251+2.607-1.900)-1)\times 100$) than nonprincelings, which equates to an advantage of 502,940.3 square meters (26,114$\times$exp$(2.251+2.607-1.900)$). Comparatively, in CK's original results (of scaled area, as reported in their paper), the presence of central inspections and Xi-appointed officials reduced the size of the princeling advantage in columns (2)-(4), per Panel A of Table \ref{tab:Table 11 dups area10k}. Thus, the reversal of this effect\textemdash and the sizable princeling advantage that persists even in the presence of the anticorruption campaign\textemdash undermines CK's conclusions about the efficacy of central inspections and Xi-appointed officials in shrinking the princeling advantage. 

These results are further echoed in Figure \ref{fig:comparison_fig6_poisson}, which again indicates that the quantity difference between princelings and nonprincelings drops prior to 2012, but the princeling advantage remains above zero from 2007-2016, with and without duplicates. When duplicates are included, the quantity difference between princelings and nonprincelings seems to decrease slightly between 2011-2016, but the 95\% confidence intervals remain wide, as Figure \ref{fig:fig6_poisson_no_duplicates} suggests. When duplicates are removed, the quantity difference between princelings and nonprincelings seems to increase slightly from 2011-2016, again suggesting the causal implications of the anticorruption campaign are limited.

\vspace{5mm}
\begin{figure}[H]
    \centering
    \caption{Recalculating CK's Figure VI, with and without duplicates, using Poisson regression}
    \begin{subfigure}[t]{0.45\linewidth}
        \centering
        \includegraphics[width=\linewidth]{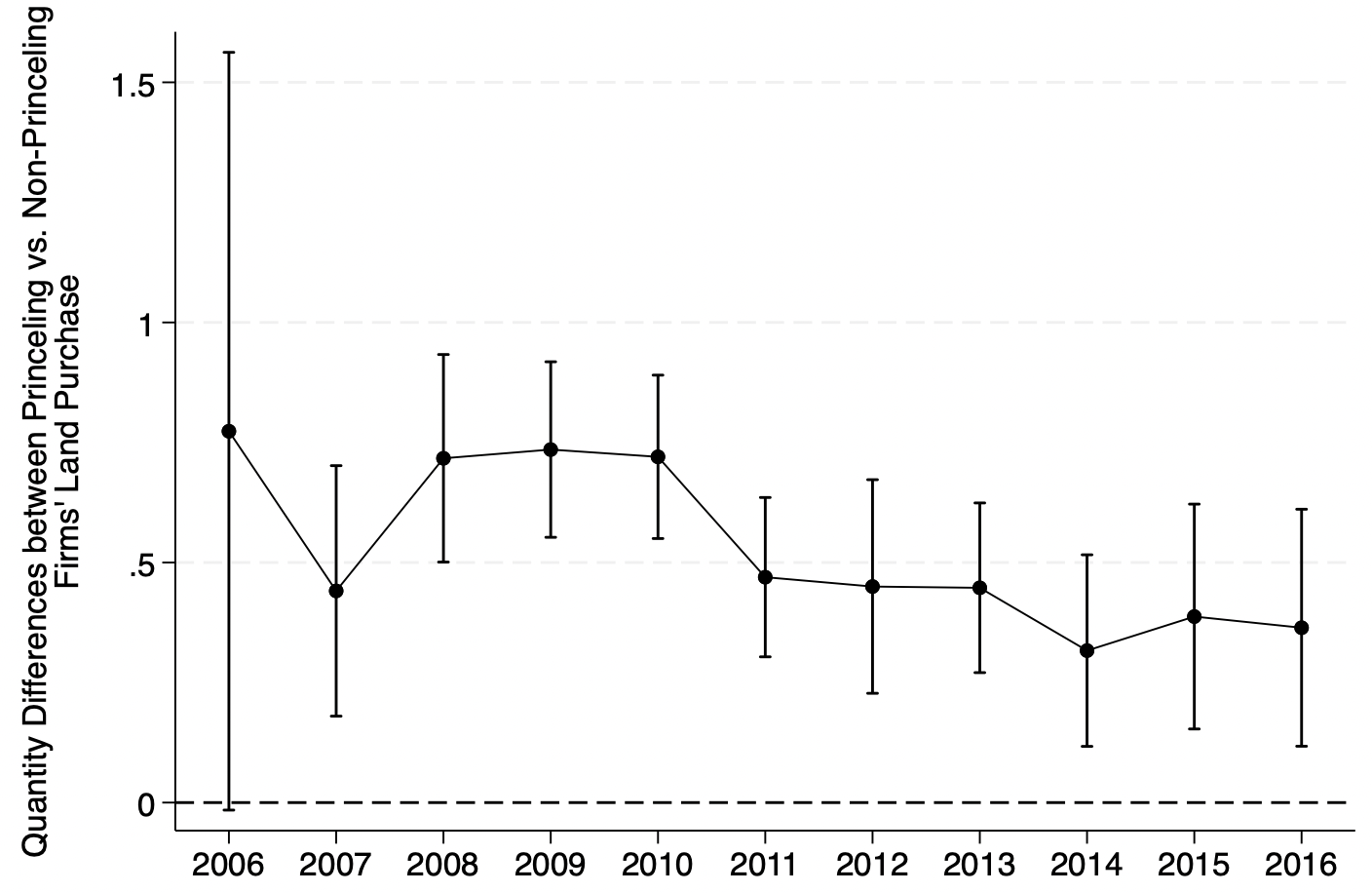}
        \caption{CK's Figure VI recalculated}
        \label{fig:fig6_poisson_with_duplicates}
    \end{subfigure}%
    \hspace{0.05\linewidth} 
    \begin{subfigure}[t]{0.45\linewidth}
        \centering
        \includegraphics[width=\linewidth]{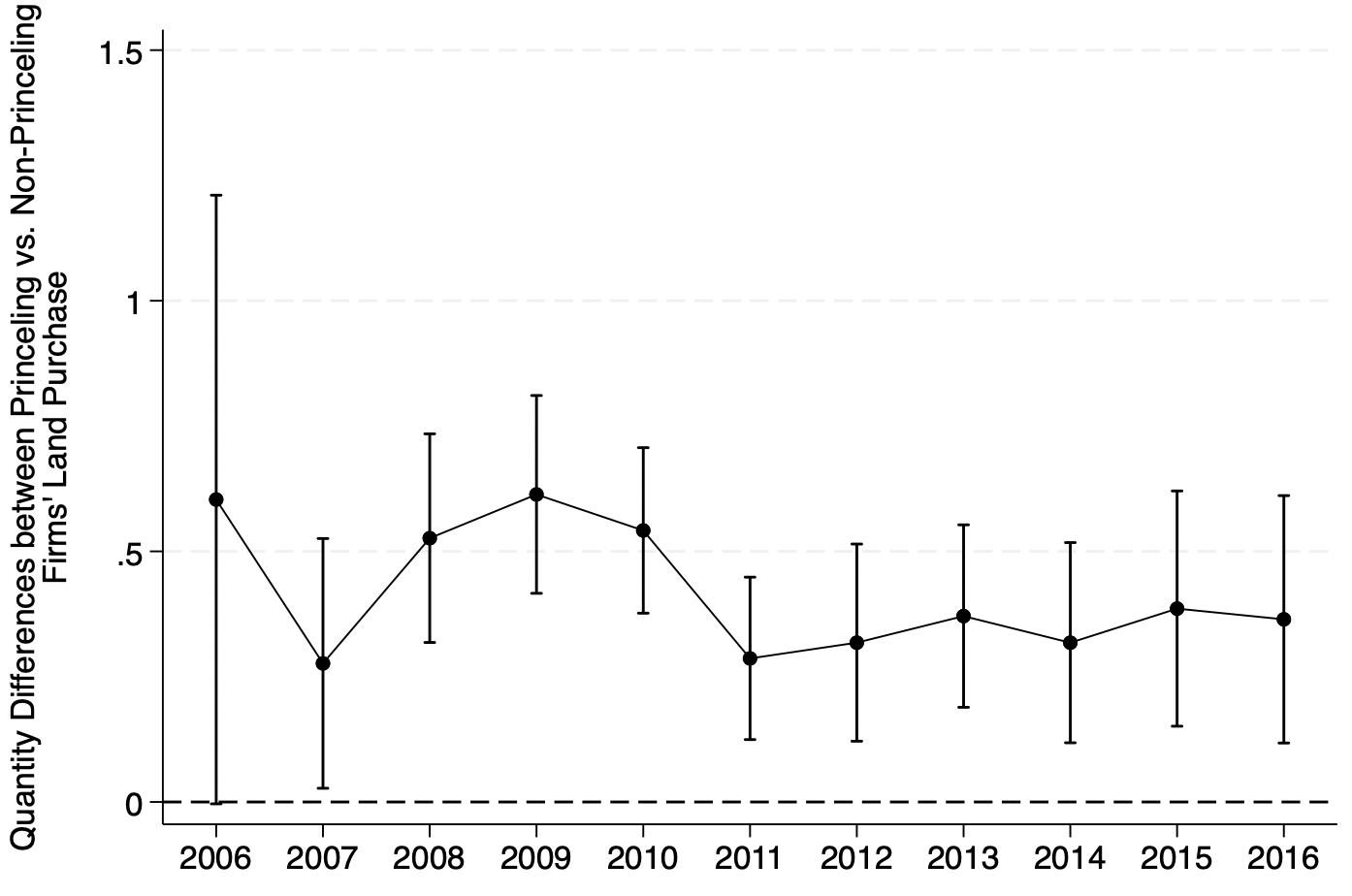}
        \caption{CK's Figure VI recalculated, no duplicates}
        \label{fig:fig6_poisson_no_duplicates}
    \end{subfigure}
    \caption*{\raggedright \footnotesize Note: Bars reflect 95\% confidence intervals.}
    \label{fig:comparison_fig6_poisson}
\end{figure}

\subsubsection{Extensive margin effect} \label{appendix: extensive margin}

I also briefly highlight the extensive margin effect, using a binary variable for any nonzero purchase area as the outcome. Recall that Table VI groups transactions by firm for each year, and Table XI groups transactions by firm in each province and year. Thus, given the dependent variable area is only binary in this check, the presence of duplicates does not change the calculation, and I therefore include results for a logistic specification and a linear probability model in each table/figure. 

\vspace{3mm}
\begin{table}[H] 
\centering
\caption{CK's Table VI, Extensive margin}
\label{tab: Table VI Appendix extensive}
\small
\begin{threeparttable}
\setlength{\tabcolsep}{6pt} 
\renewcommand{\arraystretch}{1.3} 
\begin{tabular}{l*{6}{c}}
            \hline \hline
& \multicolumn{6}{c}{Nonzero land purchased} \\
\cmidrule(lr){2-7}
                    & \multicolumn{3}{c}{Panel A: Logistic specification} & \multicolumn{3}{c}{Panel B: Linear probability model} \\
\cmidrule(lr){2-4} \cmidrule(lr){5-7}
                    & (1) & (2) & (3) & (1) & (2) & (3) \\

\midrule
Princeling firms    &       0.175\sym{***}&      -0.464\sym{***}&      -0.037\sym{*}  &       0.015\sym{***}&      -0.030\sym{***}&      -0.003\sym{*}  \\
                    &     (0.015)         &     (0.019)         &     (0.015)         &     (0.001)         &     (0.001)         &     (0.001)         \\
Princeling firms*PSCM&                     &       5.168\sym{***}&                     &                     &       0.836\sym{***}&                     \\
                    &                     &     (0.092)         &                     &                     &     (0.006)         &                     \\
Princeling firms*Retired&                     &                     &       4.738\sym{***}&                     &                     &       0.806\sym{***}\\
                    &                     &                     &     (0.147)         &                     &                     &     (0.008)         \\
\hline
\(R^{2}\)  &       0.096               &       0.099              &      0.097                &       0.044         &       0.047         &       0.045         \\
\hline\hline
\end{tabular}
\end{threeparttable}
\caption*{\footnotesize Note: Following CK, all columns include year fixed effects, as well as the control variables of firm size and state-ownership status. CK note that industry fixed effects are also included, but their code confirms that they are omitted. I follow their code and do not include industry fixed effects. Standard errors are robust to clustering by firm and are in parentheses (\sym{*} \(p<0.05\), \sym{**} \(p<0.01\), \sym{***} \(p<0.001\)). Note that the \(R^{2}\) is the pseudo \(R^{2}\) for the logistic specification and the adjusted \(R^{2}\) for the linear probability model. There are 5,690,984 observations in each column.}
\end{table}

As is apparent in Table \ref{tab: Table VI Appendix extensive}, the baseline princeling effect on this extensive margin is modest when a binary outcome is used: per column (1) of Panel A, being a princeling raises the probability of purchasing at least one land parcel by 1.39 percentage points, compared to nonprincelings.\footnote{Note that all percentage point interpretations from the logistic specification are derived via Stata's marginal effects calculation.} Column (1) of Panel B echoes this, suggesting that being a princeling increases the probability of making any land purchase by 1.5 percentage points relative to nonprincelings, holding fixed year, state-ownership status, and firm size effects. For the subsequent columns, the magnitude of the extensive margin effect is different between the logistic and linear specifications, but the direction of the coefficient remains the same across specifications. Further, the princeling effect becomes negative when interacted terms are introduced: Panel A's column (2) suggests that princelings have a 3.69 percentage point drop in the probability of purchasing at least one parcel of land, compared to nonprincelings. However, princelings with PCSM connections have a 40.99 percentage point increase in the probability of purchasing at least one parcel of land, relative to nonprincelings. For both panels, column (3) behaves similarly, yielding a small negative effect for princelings but a large positive effect when princelings have connections to retired officials. 

\begin{table}[H]
\centering
\caption{CK's Table XI, Extensive margin}
\label{tab: Table 11 Appendix extensive}
\resizebox{\textwidth}{!}{ 
\begin{threeparttable}
\setlength{\tabcolsep}{6pt} 
\renewcommand{\arraystretch}{1.3} 
\begin{tabular}{l*{8}{c}}
\hline \hline
& \multicolumn{8}{c}{Nonzero land purchased} \\
\cmidrule(lr){2-9} 
& \multicolumn{4}{c}{Panel A: Logistic specification} & \multicolumn{4}{c}{Panel B: Linear probability model} \\
\cmidrule(lr){2-5} \cmidrule(lr){6-9}
                    & (1) & (2) & (3) & (4) & (1) & (2) & (3) & (4)  \\
Princeling firms    &       4.701\sym{***}&       4.833\sym{***}&       4.851\sym{***}&       4.715\sym{***}&            0.805\sym{***}&       0.815\sym{***}&       0.814\sym{***}&       0.804\sym{***}\\
                    &     (0.130)         &     (0.128)         &     (0.133)         &     (0.135)                &     (0.008)         &     (0.007)         &     (0.008)         &     (0.008)         \\
Princeling firms &       0.192         &                     &                     &       1.084\sym{***}&                      0.009         &                     &                     &       0.064\sym{***}\\
\hspace{15pt}$*$Transactions after 2012
                    &     (0.139)         &                     &                     &     (0.245)         &                (0.008)         &                     &                     &     (0.012)         \\
Central inspection  &                     &       7.496\sym{***}&                     &       7.430\sym{***}&                              &       0.918\sym{***}&                     &       0.557\sym{***}\\
                    &                     &     (0.322)         &                     &     (0.450)         &                                   &     (0.008)         &                     &     (0.092)         \\
Princeling firms &                     &      -6.620\sym{***}&                     &      -6.953\sym{***}&                     &                -0.858\sym{***}&                     &      -0.538\sym{***}\\
\hspace{15pt}$*$Central inspection
                    &                     &     (0.279)         &                     &     (0.240)         &                                 &     (0.012)         &                     &     (0.093)         \\
Xi-appointed officials&                     &                     &       7.844\sym{***}&       7.883\sym{***}&                     &                                 &       0.934\sym{***}&       0.755\sym{***}\\
                    &                     &                     &     (0.364)         &     (0.402)         &                     &                                        &     (0.011)         &     (0.042)         \\
Princeling firms&                     &                     &      -6.635\sym{***}&      -7.237\sym{***}&                     &                                   &      -0.860\sym{***}&      -0.731\sym{***}\\
\hspace{15pt}$*$Xi-appointed officials                    &                     &                     &     (0.244)         &     (0.356)         &                     &                                       &     (0.010)         &     (0.044)         \\
\hline
Observations        &     5,758,311         &     5,758,311         &     5,758,311         &     5,758,311               &    11,516,622         &    11,516,622         &    11,516,622         &    11,516,622         \\
\(R^{2}\)  &   0.101                  &   0.194                  &     0.253                 &           0.306                &       0.096         &       0.208         &       0.269         &       0.304         \\
\hline \hline
\end{tabular}
\end{threeparttable}
}
\caption*{\footnotesize Note: Following CK, all columns include province and year fixed effects, as well as the control variables of firm size and state-ownership status. Standard errors are robust to two-way clustering by firm and province and are in parentheses (\sym{*} \(p < 0.05\), \sym{**} \(p < 0.01\), \sym{***} \(p < 0.001\)). Note that the \(R^{2}\) is the pseudo \(R^{2}\) for the logistic specification and the adjusted \(R^{2}\) for the linear probability model. 
}
\end{table}

The regressions of Table \ref{tab: Table 11 Appendix extensive} consistently yield a large princeling effect, with Panel A's column (1), for instance, suggesting that princelings have a 37.02 percentage point higher probability of purchasing at least one parcel compared to nonprincelings. This analysis on the extensive margin maintains a strong princeling effect despite all of CK's interacted variables, with the princeling effect only increasing after 2012 (Panel A column (1), by a non-statistically significant 1.5 percentage points) and increasing further in the presence of central inspections. The linear probability model echoes these findings, with Panel B's column (1) suggesting that being a princeling increases the probability of purchasing any land by 80.5 percentage points relative to nonprincelings, holding fixed province, year, state-ownership status, and firm size effects. Panel B's column (3), for example, suggests that for transactions occurring in provinces wherein Xi replaced the party secretary, princelings had an 88.8 percentage point higher $((0.814 + 0.934 - 0.860)\times$100) probability of purchasing at least one parcel compared to nonprincelings. Thus, in this analysis of the extensive margin effect, the anticorruption campaign's central inspections and Xi-appointed officials do not shrink the probability of princelings purchasing at least one parcel. 

Conversely, Figure \ref{fig:comparison_fig6_extensive} shows a shrunk princeling effect on the extensive margin for both methods, with princelings being less likely to purchase land compared to nonprincelings in 2014-2015. However, the difference in purchase probabilities begins to decline in 2011 prior to the anticorruption campaign, with the 95\% confidence interval including zero in 2012 and hovering at or below zero through 2016. 

\vspace{5mm}
\begin{figure}[H]
    \centering
    \caption{Recalculating CK's Figure VI, Extensive margin}
    \begin{subfigure}[t]{0.5\linewidth}
        \centering
        \includegraphics[width=\linewidth]{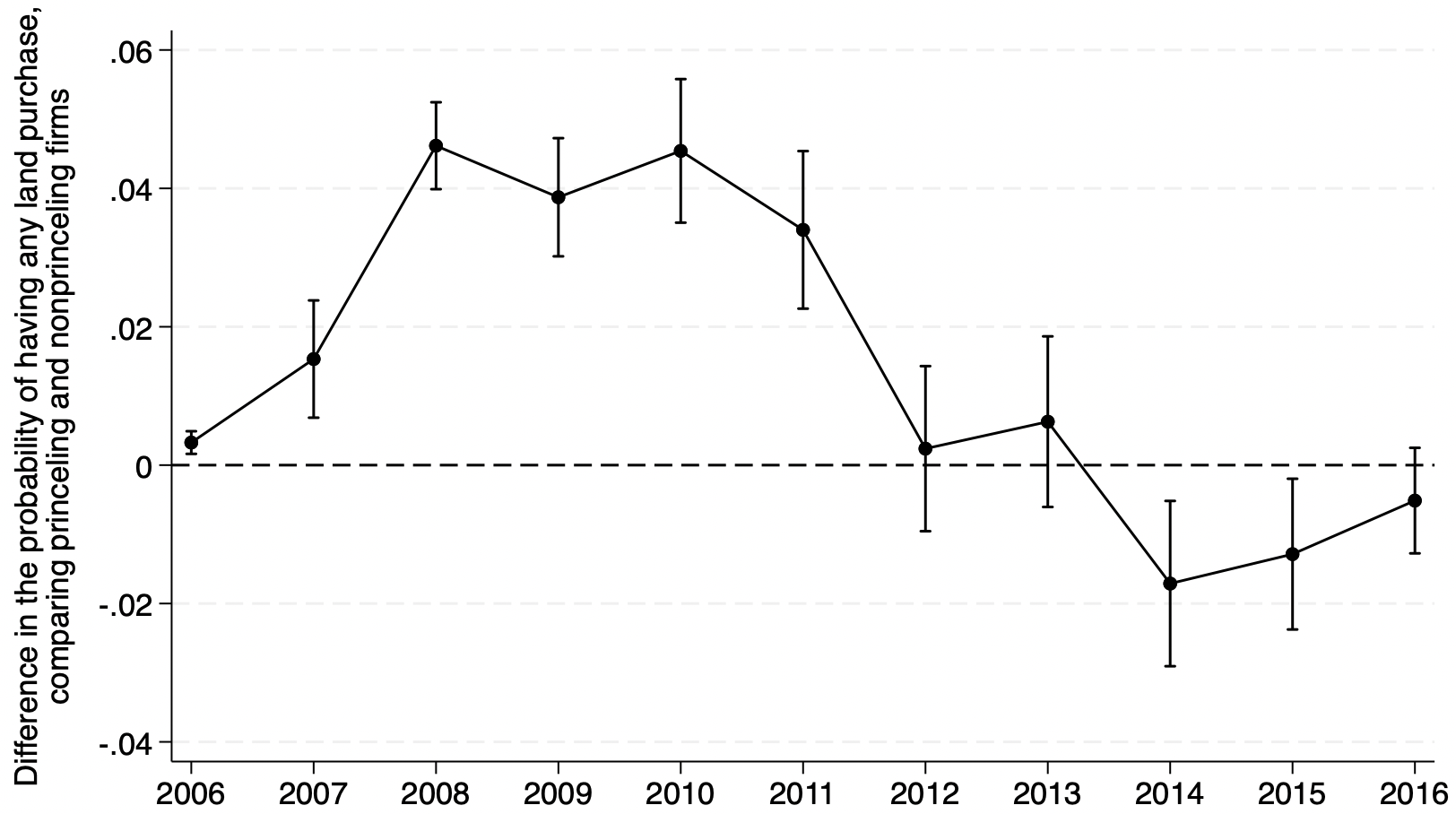}
        \caption{CK's Figure VI recalculated, logistic}
        \label{fig:fig6_with_duplicates_extensive}
    \end{subfigure}%
    \hspace{0.04\linewidth} 
    \begin{subfigure}[t]{0.45\linewidth}
        \centering
        \includegraphics[width=\linewidth]{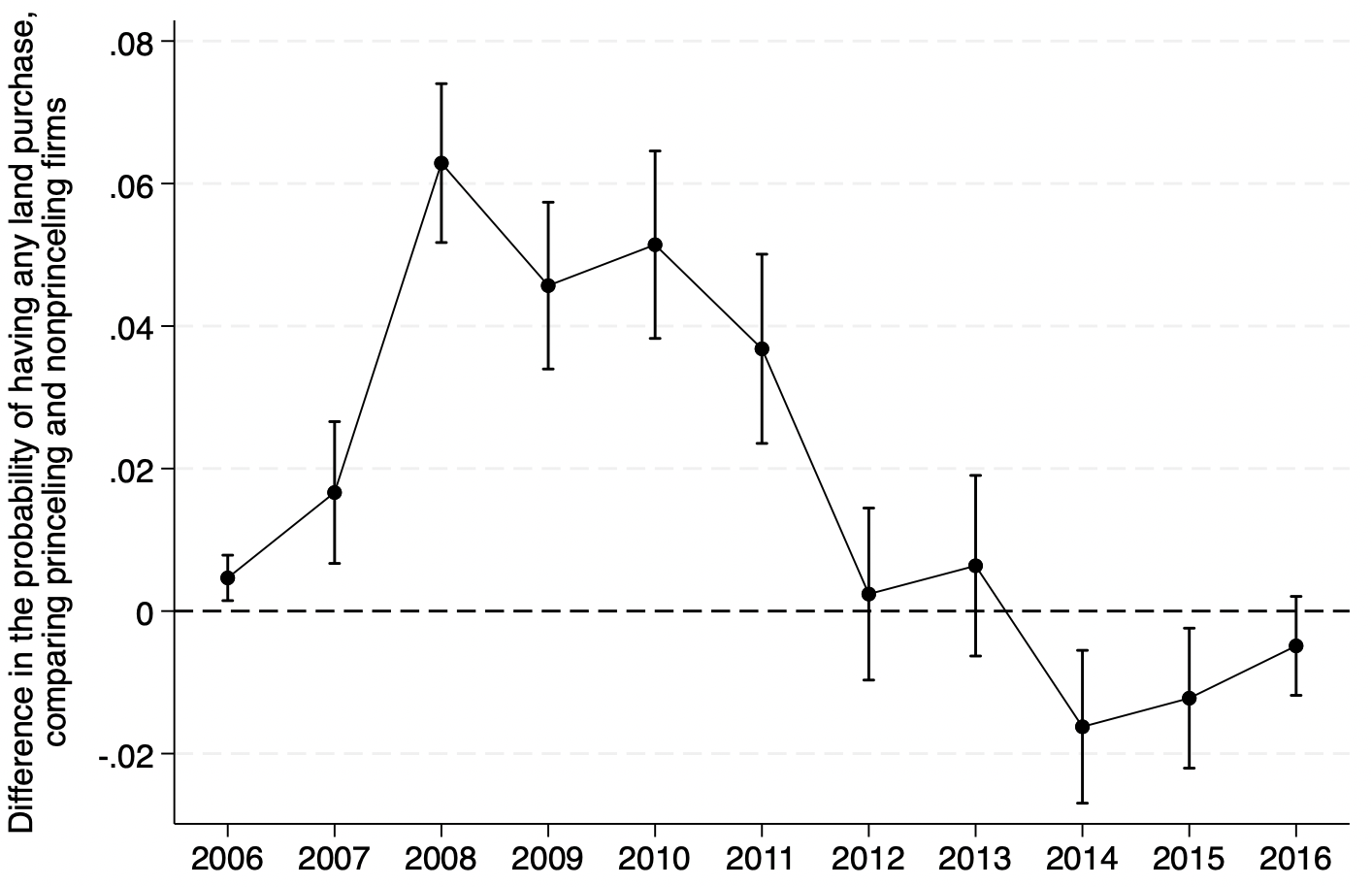}
        \caption{CK's Figure VI recalculated, linear}
        \label{fig:fig6_no_duplicates_extensive}
    \end{subfigure}
    \caption*{\raggedright \footnotesize Note: Bars reflect 95\% confidence intervals. Note also that Panel A plots the marginal effects of the princeling firm variable evaluated at each year. }
    \label{fig:comparison_fig6_extensive}
\end{figure}

Ultimately, analyzing these results in tandem with the baseline Poisson specification reveals that princeling effects are, on average, smaller on the extensive margin, although their sign and direction vary across specifications. In some cases, the princeling effect reverses sign relative to the baseline Poisson specifications reported in Table \ref{tab: Table VI Appendix poisson}. This heterogeneity suggests that the intensive and extensive margins are sometimes aligned and sometimes diverging. However, despite these complexities, these results provide little support for CK's original findings about the mechanisms and efficacy of the anticorruption campaign.
\end{document}